\documentclass[usenatbib]{texstuff/mn2e}
\usepackage{graphicx}
\usepackage{xcolor}
\usepackage{amssymb,amsmath}
\usepackage{natbib}
\usepackage{rotating,pdflscape}
\usepackage{adjustbox}
\usepackage{longtable}

\newcommand{\ha}{\mbox{H$\alpha$}} 
\newcommand{\nii}{[N{\sc II}]}

\newcommand{\GS}{$\Gamma$-$\Sigma_{\rm SFR}$} 
\newcommand{\sfrsig}{$\Sigma_{\rm SFR}$} 
\newcommand{\mvsfr}{$M_{\rm V}^{\rm brightest}$-SFR} 
\newcommand{\mv}{$M_{\rm V}^{\rm brightest}$} 
\newcommand{\aper}{D25$\cap$HST}
\newcommand{\sfrunit}{\rm{M}_{\odot}\rm{yr}^{-1}}
\newcommand{\sigunit}{\rm{M}_{\odot}\rm{yr}^{-1}\rm{kpc}^{-2}}

\newcommand{\Glow}{$\Gamma_{1-10}$}
\newcommand{\Gall}{$\Gamma_{1-100}$}
\newcommand{\Gmid}{$\Gamma_{10-100}$}

\begin{document}

\title{Fraction of Stars in Clusters for the LEGUS Dwarf Galaxies}

\author[D.O. Cook et al.]
{D.O. Cook$^{1}$,
J.C. Lee$^{2,3}$,
A. Adamo$^{4}$,
D. Calzetti$^{5}$,
R. Chandar$^{6}$,
B.C. Whitmore$^{7}$,
\newauthor
A. Aloisi$^{7}$,
M. Cignoni$^{8,9,10}$,
D.A. Dale$^{11}$,
B.G. Elmegreen$^{12}$,
\newauthor
M. Fumagalli$^{13,10}$,
K. Grasha$^{14}$,
K.E. Johnson$^{15}$,
R.C. Kennicutt$^{16,17}$,
H. Kim$^{18}$,
\newauthor
S.T. Linden$^{5,15}$,
M. Messa$^{4,19}$,
G. \"Ostlin$^{4}$,
J.E. Ryon$^{7}$,
E. Sacchi$^{10,20,7}$,
\newauthor
D.A. Thilker$^{21}$,
M. Tosi$^{10}$,
A. Wofford$^{22}$
\\ 
$^{1}$Caltech/IPAC, 1200 E. California Boulevard, Pasadena, CA 91125, USA\\ 
$^{2}$Dept. of Astronomy, University of Arizona, Tucson, AZ\\ 
$^{3}$Gemini Observatory/NOIRLab, Tucson, AZ 85719, USA\\ 
$^{4}$Dept. of Astronomy, The Oskar Klein Centre, Stockholm University, Stockholm, Sweden\\ 
$^{5}$Dept. of Astronomy, University of Massachusetts -- Amherst, Amherst, MA\\ 
$^{6}$Dept. of Physics and Astronomy, University of Toledo, Toledo, OH\\ 
$^{7}$Space Telescope Science Institute, Baltimore, MD\\ 
$^{8}$Department of Physics, University of Pisa, Largo B. Pontecorvo 3, 56127, Pisa, Italy\\ 
$^{9}$INFN, Largo B. Pontecorvo 3, 56127, Pisa, Italy\\ 
$^{10}$INAF -- OAS Osservatorio di Astrofisica e Scienza dello Spazio, Bologna, Italy\\ 
$^{11}$Dept. of Physics \& Astronomy, University of Wyoming, Laramie, WY\\ 
$^{12}$IBM Research Division, T.J. Watson Research Center, Yorktown Hts., NY\\ 
$^{13}$Dipartimento di Fisica G. Occhialini, Universit\`a degli Studi di Milano Bicocca, Piazza della Scienza 3, 20126 Milano, Italy\\ 
$^{14}$Research School of Astronomy and Astrophysics, Australian National University, Canberra, Australia\\ 
$^{15}$Dept. of Astronomy, University of Virginia, Charlottesville, VA\\ 
$^{16}$Steward Observatory, University of Arizona, Tucson, AZ\\ 
$^{17}$George P. and Cynthia W. Mitchell Institute for Fundamental Physics \& Astronomy, Texas A\&M University, College Station, TX\\ 
$^{18}$Gemini Observatory, Casilla 603, La Serena, Chile\\ 
$^{19}$Observatoire de Gen\`eve, University of Geneva, Geneva,Switzerland\\ 
$^{20}$Dept. of Physics and Astronomy, Bologna University, Bologna, Italy\\ 
$^{21}$Dept. of Physics and Astronomy, The Johns Hopkins University, Baltimore, MD\\ 
$^{22}$Instituto de Astronomia, Universidad Nacional Autonoma de Mexico, Unidad Acad\'emica en Ensenada, Ensenada, Mexico\\ 
}

\maketitle

\begin{abstract}

We study the young star cluster populations in 23 dwarf and irregular galaxies observed by the HST Legacy ExtraGalactic Ultraviolet Survey (LEGUS), and examine relationships between the ensemble properties of the cluster populations and those of their host galaxies: star formation rate (SFR) density (\sfrsig).  A strength of this analysis is the availability of SFRs measured from temporally resolved star formation histories which provide the means to match cluster and host-galaxy properties on several timescales (1-10, 1-100, and 10-100~Myr). Nevertheless, studies of this kind are challenging for dwarf galaxies due to the small numbers of clusters in each system.  We mitigate these issues by combining the clusters across different galaxies with similar \sfrsig\ properties. We find good agreement with a well-established relationship ($M_{V}^{brightest}$--SFR), but find no significant correlations between \sfrsig\ and the slopes of the cluster luminosity function, mass function, nor the age distribution. We also find no significant trend between the the fraction of stars in bound clusters at different age ranges (\Glow, \Gmid, and \Gall) and \sfrsig\ of the host galaxy. Our data show a decrease in $\Gamma$ over time (from 1-10 to 10-100~Myr) suggesting early cluster dissolution, though the presence of unbound clusters in the youngest time bin makes it difficult to quantify the degree of dissolution. While our data do not exhibit strong correlations between \sfrsig\ and ensemble cluster properties, we cannot rule out that a weak trend might exist given the relatively large uncertainties due to low number statistics and the limited \sfrsig\ range probed. 

\end{abstract}


\clearpage

\section{INTRODUCTION} 

What are the conditions that form long-lived, bound star clusters which may survive to be globular clusters in the present day universe?  Such conditions seem to be common in the past, but not in the nearby galaxy population as globular cluster progenitors seem to be rare today.  Interestingly however, super star clusters ($M>10^5~{\rm M}_{\odot}$) have been observed in dwarf starburst galaxies \citep[e.g., NGC~1705, N5253;][]{billett02,vasquez04,martins12,calzetti15,turner15,Lsmith20}. Compared to larger spirals and massive interacting galaxies, much less is known about the cluster populations of low-mass galaxies (i.e., dwarfs and irregulars) where their relatively low SFRs produce proportionally fewer clusters \citep{whitmore00,larsen02,bastian08}. A better understanding of the cluster populations in dwarf galaxies is an important step toward understanding star formation and the extreme environments of the past. 

High resolution observations of dwarf galaxies enable the study of conditions in the low-mass and low-star formation density regime, which can provide strong constraints on the relationships between cluster populations and their host-galaxy properties. Such observational relationships are key inputs for theoretical studies that try to model the interplay between the gaseous natal material and various feedback mechanisms that govern the star formation process \citep[e.g.,][]{kruijssen12,Hopkins13a,Guszejnov17,Kim18}. 

There are two relationships reported in the literature that connect the properties of cluster populations to those of their host galaxies: 1) The brightest cluster versus the galaxy star formation rate (\mvsfr).  2) The fraction of stars found in bound clusters ($\Gamma$) versus the SFR density (\GS).  These relationships can reveal whether more clusters are retained/formed under particular conditions and may provide the means to better understand how clusters formed in the past. These relationships, and possibly others, provide key insights into the star formation process. 

Other properties of the ensemble cluster population have also been useful to compare the cluster formation and dissolution rates between different galaxies. The cluster mass (luminosity) function is often described as a power law of the form $dN/dM \propto M^{\beta}$ ($dN/dL \propto L^{\alpha}$) where most cluster populations exhibit a $\beta=-2$ and $\alpha=-2$ index \citep{kennicutt12}, which is consistent with a scale-free star formation process \citep{elmegreen10}. Additionally, the age distribution of cluster populations, $dN/dt \propto t^{\gamma}$, provides a quantitative measure of both cluster formation and dissolution over time. At this time it is not clear if these fundamental cluster properties correlate with galaxy-wide environment \citep[][]{weidner04,bastian12a,Randriamanakoto13,Pflamm13,silvavilla14,whitmore14a,adamo15,cook16,johnson17,messa18b,messa18a,cook19a,Santoro22}.

When investigating relationships between cluster populations and their host galaxies, it is also critical to take into account systematic effects that can act to wash out (reduce) the strength of these relationships. In particular, it is important to perform analysis over different ranges of star cluster ages, and ensure that cluster properties and those of the host galaxy are computed self-consistently over matched timescales. This is important because cluster mass is lost during the process of star cluster dissolution over time. While the exact mechanisms of cluster dissolution is under debate \citep[for a discussion of internal and external processes, see][]{krumholz19}, it is widely accepted that star clusters will dissolve as can be seen in the age distributions of clusters in nearby galaxies \citep[e.g.,][]{fall12}. The timescales of cluster dissolution can be quantified in the nearest, resolved star clusters, which have crossing times of $\approx$10~Myr \citep{gieleszwart11}. Consequently, clusters that have an age less than (several) crossing times may look centrally concentrated but may also be in the process of dissolving, and can contaminate catalogs of bound clusters. The recent study of \cite{Brown21} has compared the instantaneous crossing times and ages of individual star clusters in 31 LEGUS galaxies suggesting that while most clusters in the sample are likely gravitationally bound, those in the 1-10~Myr age range tend to have a lower bound fraction ($f_{\rm bound}\approx20-70\%$), as may be expected.

In this study we utilize the data products from the LEGUS HST survey \citep{calzetti15}. LEGUS includes galaxies with a wide range of global star formation properties, and has yielded star formation histories with high temporal resolution \citep{Cignoni18,sacchi18,cignoni19}, and robust star cluster catalogs \citep{adamo17,cook19a}. Moreover, half of the LEGUS sample is comprised of dwarf and irregular galaxies, which allows an order of magnitude increase of the aggregate star cluster sample size compared to previous studies \citep{cook12}. With these data we can provide more stringent constraints on several relationships between cluster and host galaxy properties. 

The outline of the remainder of this paper is as follows.  In Section 2, we describe the properties of galaxies in this analysis, with particular focus on the derivation of star formation rates.  In Section 3, we describe the LEGUS star cluster catalogs used.  In Section 4, the coverage of the galaxies is discussed.  Results are presented in Section 5, and include analysis of the relationship between the magnitude of the brightest cluster and the SFR of the host galaxy; star cluster luminosity and mass functions; star cluster age distributions (dN/dt); and the fraction of stars in clusters ($\Gamma$).  In this paper we define $\Gamma$ as the fraction of stellar mass located in bound star clusters \citep{bastian08} as measured in three age ranges: \Glow, \Gmid, and \Gall\ corresponding to age ranges of 1-10, 10-100, and 1-100~Myr, respectively. In Section 6, the results are discussed in the context of previous results and remaining uncertainties which should be addressed by future work.  We conclude in Section 7 with a summary of the results. Note that all logarithmic scales used in this paper correspond to log-base 10. In addition, the mathematical symbols used in this paper are defined as follows: $t$, $M$, and $N$ refer to age, mass, and number of clusters, respectively.

\section{Galaxy Properties}

\subsection{Sample}

The data used in this analysis come from the HST Legacy ExtraGalactic UV Survey \citep[LEGUS;][]{legus}, from which we have chosen a sub-sample of dwarf and irregular galaxies. A full description of this sub-sample is provided in \citet[][hereafter C19]{cook19a}, but we provide a brief overview of the selection criteria and properties here. 

The LEGUS dwarf and irregular galaxies were chosen based on the absence of obvious spiral arms and dust lanes in the HST color images. As a result of this morphological selection, some of the galaxies in this sub-sample have irregular morphologies and may not strictly be considered dwarf galaxies by their stellar mass. However, the majority (all but two, NGC~4449 and NGC~4656) have stellar masses below log($M_{\star} [\rm{M}_{\odot}])\leq9$. There are 23 (out of 50) galaxies in the LEGUS sample that meet these morphological criteria (Table~\ref{tab:globalprop}). We note that the cluster catalogs for 17 of these galaxies were available at the time of the C19 study. However, the catalogs for all 23 galaxies have since been completed and we utilize the full sample in this analysis. 

Examination of the physical properties of this morphologically selected sample confirm that the galaxies tend to have low SFRs, low stellar masses (M$_{\star}$), and low metallicities as expected for dwarf galaxies; these properties are presented in Table~1 of C19. As discussed there, the global properties of this sub-sample span a range in metallicity of 0.001 $\leq$ Z $\leq$0.02, SFR of $-2.30<\rm{log(SFR [\sfrunit])<-0.03}$, stellar mass of $7.3<\rm{log(M_{\star} [M_{\odot}]) <9.5}$, and SFR surface density of  $-3.1<\rm{log(\Sigma_{\rm{SFR}} [\sigunit])}<-1.5$ (see C19 for the methods used to derive these properties). However, we note that these SFRs and SFR surface densities are not the values used later in this study since they are averaged over the full extent of the galaxy, and may not accurately reflect the activity of the area covered by the LEGUS HST footprint (see \S\ref{sec:areatest}).

In addition, LEGUS tip of the red giant branch (TRGB) distance measurements for the galaxies have been published by \cite{sabbi18} since the compilation in \cite{legus}. We adopt the \cite{sabbi18} distances for this analysis, and have scaled the cluster masses to account for any changes in adopted distance (since the distances in \cite{legus} were assumed when generating the cluster catalogs). We note that the star formation histories (SFHs) derived from resolved stellar population color-magnitude diagrams for the LEGUS dwarfs \citep{cignoni19} assumed these same distances (provided in Table~\ref{tab:globalprop}). The majority of the galaxy distances did not change significantly with a median change of 2\%. However, UGC 695 has an updated distance that is 3.1~Mpc closer, and three galaxies (IC 559, NGC 3274, and NGC 4656) are several Mpc further away. The resulting change in cluster mass for most galaxies is less than 0.1~dex, but can be as high as 0.5~dex for the galaxies with the largest change in distances.

\begin{table*}

{Global Galaxy Properties}\\
\begin{tabular}{lccccc}

\hline
\hline
&&& Dist & Dist \\
Galaxy         & RA       & DEC      & Old   & New  \\ 
Name           & (J2000)  & (J2000)  & (Mpc)     & (Mpc)    \\ 
\hline

UGC00685 & $01:07:22.4$ & $+16:41:04$    & ~4.8 & ~4.4 $\pm$ 0.3 \\
UGC00695 & $01:07:46.4$ & $+01:03:49$    & 10.9 & ~7.8 $\pm$ 0.6 \\
UGC01249 & $01:47:29.9$ & $+27:20:00$    & ~6.9 & ~6.4 $\pm$ 0.5 \\
NGC1705 & $04:54:13.5$ & $-53:21:40$     & ~5.1 & ~5.2 $\pm$ 0.4 \\
ESO486-G021 & $05:03:19.7$ & $-25:25:23$ & ~9.5 & ~9.1 $\pm$ 0.7 \\
UGC04305 & $08:19:05.0$ & $+70:43:12$    & ~3.0 & ~3.3 $\pm$ 0.3 \\
UGC04459 & $08:34:07.2$ & $+66:10:54$    & ~3.7 & ~4.0 $\pm$ 0.3 \\
UGC05139 & $09:40:35.1$ & $+71:10:46$    & ~4.0 & ~3.8 $\pm$ 0.3 \\
IC0559 & $09:44:43.9$ & $+09:36:54$      & ~5.3 & 10.0 $\pm$ 0.9 \\
UGC05340 & $09:56:45.7$ & $+28:49:35$    & 12.7 & 12.7 $\pm$ 1.0 \\
NGC3274 & $10:32:17.3$ & $+27:40:08$     & ~6.6 & 10.0 $\pm$ 0.9 \\
NGC3738 & $11:35:48.8$ & $+54:31:26$     & ~4.9 & ~5.1 $\pm$ 0.4 \\
UGC07242 & $12:14:08.4$ & $+66:05:41$    & ~5.4 & ~5.7 $\pm$ 0.4 \\
NGC4248 & $12:17:49.8$ & $+47:24:33$     & ~7.8 & ~6.8 $\pm$ 0.5 \\
UGC07408 & $12:21:15.0$ & $+45:48:41$    & ~6.7 & ~7.0 $\pm$ 0.5 \\
UGCA281 & $12:26:15.9$ & $+48:29:37$     & ~5.9 & ~5.2 $\pm$ 0.4 \\
NGC4449 & $12:28:11.1$ & $+44:05:37$     & ~4.3 & ~4.0 $\pm$ 0.3 \\
NGC4485 & $12:30:31.1$ & $+41:42:04$     & ~7.6 & ~8.8 $\pm$ 0.7 \\
NGC4656 & $12:43:57.7$ & $+32:10:05$     & ~5.5 & ~7.9 $\pm$ 0.7 \\
IC4247 & $13:26:44.4$ & $-30:21:45$      & ~5.1 & ~5.1 $\pm$ 0.4 \\
NGC5238 & $13:34:42.5$ & $+51:36:49$     & ~4.5 & ~4.4 $\pm$ 0.3 \\
NGC5253 & $13:39:56.0$ & $-31:38:24$     & ~3.2 & ~3.3 $\pm$ 0.3 \\
NGC5477 & $14:05:33.3$ & $+54:27:40$     & ~6.4 & ~6.7 $\pm$ 0.5 \\

\hline
\end{tabular}
\caption{Properties of the LEGUS dwarf galaxies. Column 1: Galaxy name. Column 2 and 3: J2000 right ascension and declination from NED. Column 4: Distances reported in \citet{legus}.  Column 5: Distance from \citet{sabbi18} derived via the TRGB method.   }

\label{tab:globalprop}

\end{table*}

\subsection{Star Formation Rates}  \label{sec:sfrs}

In this section we describe the various galaxy-wide star formation rate (SFR) indicators that are utilized to examine whether correlations exist between the ensemble properties of the star cluster populations and the properties of their host galaxies. With the high resolution HST data for the LEGUS galaxies we are able to derive SFRs averaged over different age ranges using SFHs based upon modeling of stellar color-magnitude diagrams (CMDs).  We also compute SFRs from the integrated \ha, UV, and IR luminosities using standard recipes. A comparison of these various SFRs are presented at the end of this section.  

\subsubsection{SFRs: Resolved Stellar Population Measurements} \label{sec:sfhs}

When comparing cluster population and host-galaxy properties, it is important to match the timescales over which these properties are computed to separate effects due to variation in star formation history. An advantage of our study is that we are able to derive both the overall SFR activity using CMD-fitting methods, and the cluster formation rates over consistent age ranges. In this section, we describe the total SFR based on the combined formation rate of resolved stars (i.e., the SFH), clusters, and stellar associations; hereafter referred to as the ``total stellar population SFR."

A galaxy's SFH, the SFR as a function of time, is measured by modeling the CMD of the individually resolved stars based on HST imaging \citep[$D < 15~$Mpc;][]{weisz08,Tolstoy09,mcquinn11,villaLarsen11,mcquinn15,Annibali22}, where impressive temporal resolution ($\Delta t/t \approx 40-50\%$) at ages $<$ 100~Myr can be obtained. The age ranges previously examined in studies of cluster-host galaxy relationships is generally at ages$<100$~Myr (often 1-10 or 1-100~Myr; due to sensitivity of integrated-light SFR indicators such as the H$\alpha$ and UV flux to these age ranges). Consequently, the corresponding cluster populations of these previous studies likely include contamination from young (1-10~Myr), unbound associations \citep{gieleszwart11}. A main goal of this study is to utilize the total stellar population SFRs to provide a more robust analysis by focusing on the 10-100~Myr formation rates of stars and clusters \citep{villaLarsen11,adamo15,johnson16,messa18a,Randriamanakoto19}.  

In this study, we use the SFHs published by \cite{cignoni19} based on CMD modeling of resolved stars for all 23 LEGUS galaxies studied here. The following assumptions are made to derive the SFHs: the PARSEC-COLIBRI stellar population models with static stellar rotation \citep{bressan12,marigo17}, a \cite{kroupa01} IMF with a mass range of 0.1 -- 300~$M_{\odot}$, a \cite{cardelli89} extinction law, TRGB distances as measured by \cite{sabbi18}, metallicity is an increasing function of time, and the youngest age stars are fixed to the gas metallicity values in \cite{calzetti15}. We note here that the methods of \cite{cignoni19} account for stochasticity in the presence of young massive stars and account for incompleteness.


While the SFHs provide the formation rate of the population of individually resolved stars, the total SFR still needs to be computed by including the formation rate of star clusters and associations of similar age ranges to those of the SFHs: 1-10~Myr, 1-100~Myr, and 10-100~Myr. This step is required since only point sources are used in the construction of the observed CMD (i.e., excluding extended sources such as clusters and associations). Thus, we add the mass of clusters and associations (class 1-3; see \S\ref{sec:clustcat}) to those derived in the SFHs to obtain the total stellar population SFR in the appropriate age ranges. We note that the cluster/association mass has been derived using similar assumptions of IMF, metallicity, etc (see \S\ref{sec:clustcat}), and has been corrected for incompleteness given a cluster mass function in each age range (see \S\ref{sec:gamma}). Table~\ref{tab:sfh} provides the total SFRs over several age ranges.

\begin{table*}

{Star Formation Rates Derived from SFHs}\\
\begin{tabular}{lcccccc}

\hline
\hline
Galaxy         & SFR                     & Fraction & SFR                    & Fraction & SFR                     & Fraction  \\ 
Name           & (1--10~Myr)             & in SFH   & (1--100~Myr)           & in SFH   & (10--100~Myr)           & in SFH    \\ 
               & (M$_{\odot}$ yr$^{-1}$) & (\#)     & (M$_{\odot}$ yr$^{-1}$)& (\#)     & (M$_{\odot}$ yr$^{-1}$) & (\#)      \\ 
\hline

UGC7408    & 1.34 $\pm$ 0.66E$-02$ & 0.00 & 5.75 $\pm$ 0.76E$-03$ & 0.00 & 5.23 $\pm$ 0.51E$-03$ & 0.00 \\
UGC5139    & 7.17 $\pm$ 0.98E$-03$ & 0.56 & 9.95 $\pm$ 0.57E$-03$ & 0.97 & 1.06 $\pm$ 0.06E$-02$ & 0.97 \\
UGC4459    & 9.62 $\pm$ 1.42E$-03$ & 0.67 & 3.48 $\pm$ 0.37E$-03$ & 0.92 & 3.15 $\pm$ 0.38E$-03$ & 0.90 \\
UGC4305    & 1.43 $\pm$ 0.16E$-02$ & 0.78 & 1.55 $\pm$ 0.06E$-02$ & 0.96 & 1.60 $\pm$ 0.06E$-02$ & 0.96 \\
NGC5238    & 1.61 $\pm$ 0.14E$-02$ & 0.37 & 8.60 $\pm$ 0.65E$-03$ & 0.88 & 8.07 $\pm$ 0.70E$-03$ & 0.95 \\
UGC5340    & 3.94 $\pm$ 0.56E$-02$ & 0.42 & 6.77 $\pm$ 0.81E$-02$ & 0.84 & 7.16 $\pm$ 0.90E$-02$ & 0.85 \\
ESO486G021 & 2.22 $\pm$ 1.73E$-02$ & 0.62 & 1.72 $\pm$ 0.25E$-02$ & 0.90 & 1.71 $\pm$ 0.21E$-02$ & 0.91 \\
UGC7242    & 6.71 $\pm$ 0.87E$-03$ & 0.53 & 8.04 $\pm$ 0.90E$-03$ & 0.96 & 8.52 $\pm$ 1.00E$-03$ & 0.96 \\
IC559      & 2.06 $\pm$ 0.16E$-02$ & 0.49 & 3.72 $\pm$ 1.09E$-02$ & 0.92 & 3.92 $\pm$ 1.21E$-02$ & 0.94 \\
NGC5477    & 2.63 $\pm$ 0.56E$-02$ & 0.85 & 2.49 $\pm$ 0.41E$-02$ & 0.99 & 2.51 $\pm$ 0.45E$-02$ & 0.99 \\
NGC4248    & 3.92 $\pm$ 3.08E$-02$ & 0.13 & 3.94 $\pm$ 3.48E$-02$ & 0.83 & 3.98 $\pm$ 3.85E$-02$ & 0.90 \\
UGC685     & 1.01 $\pm$ 1.25E$-02$ & 0.21 & 6.35 $\pm$ 1.26E$-03$ & 0.89 & 6.32 $\pm$ 0.60E$-03$ & 0.95 \\
NGC1705    & 1.52 $\pm$ 0.98E$-01$ & 0.33 & 4.13 $\pm$ 0.98E$-02$ & 0.63 & 3.01 $\pm$ 0.45E$-02$ & 0.78 \\
UGCA281    & 1.20 $\pm$ 0.14E$-02$ & 0.69 & 6.92 $\pm$ 1.05E$-03$ & 0.95 & 6.71 $\pm$ 1.15E$-03$ & 0.95 \\
IC4247     & 7.33 $\pm$ 2.48E$-03$ & 0.41 & 9.12 $\pm$ 0.85E$-03$ & 0.87 & 9.50 $\pm$ 0.91E$-03$ & 0.89 \\
UGC695     & 2.07 $\pm$ 0.71E$-02$ & 0.75 & 3.91 $\pm$ 2.37E$-02$ & 0.97 & 4.13 $\pm$ 2.63E$-02$ & 0.98 \\
UGC1249    & 1.02 $\pm$ 0.06E$-01$ & 0.91 & 2.46 $\pm$ 0.71E$-01$ & 0.98 & 2.63 $\pm$ 0.78E$-01$ & 0.98 \\
NGC3274    & 2.15 $\pm$ 0.14E$-01$ & 0.89 & 1.93 $\pm$ 0.92E$-01$ & 0.97 & 1.91 $\pm$ 1.02E$-01$ & 0.97 \\
NGC5253    & 2.21 $\pm$ 0.27E$-01$ & 0.60 & 8.90 $\pm$ 0.63E$-02$ & 0.83 & 7.55 $\pm$ 0.64E$-02$ & 0.89 \\
NGC4485    & 3.53 $\pm$ 0.69E$-01$ & 0.68 & 3.16 $\pm$ 0.45E$-01$ & 0.82 & 3.14 $\pm$ 0.50E$-01$ & 0.83 \\
NGC3738    & 1.13 $\pm$ 0.13E$-01$ & 0.75 & 1.67 $\pm$ 0.10E$-01$ & 0.91 & 1.74 $\pm$ 0.11E$-01$ & 0.92 \\
NGC4449    & 6.52 $\pm$ 3.50E$-01$ & 0.60 & 4.77 $\pm$ 0.35E$-01$ & 0.91 & 4.65 $\pm$ 0.16E$-01$ & 0.94 \\
NGC4656    & 4.67 $\pm$ 0.16E$-01$ & 0.70 & 1.78 $\pm$ 0.25E$+00$ & 0.96 & 1.93 $\pm$ 0.28E$+00$ & 0.97 \\
\hline

\end{tabular}

\caption{The total stellar population SFRs for the LEGUS dwarf galaxy sample in three age ranges. The total SFRs are derived from a combination of SFHs taken from \citet{cignoni19} and the formation rates of clusters and associations that are not included in the SFH-based measurements. Columns to the right of the SFRs in each age range indicate the fraction of the total SFR coming from the resolved stars (i.e., SFHs). The galaxies are sorted by \sfrsig~using the 10-100~Myr SFHs and the \aper~areas (see \S\ref{sec:areatest}).  We note that the SFRs for UGC 7408 are lower limits as no measurable SFR could be derived from CMD fitting. }
\label{tab:sfh}

\end{table*}

\subsubsection{SFRs: Integrated-light Measurements} \label{sec:ilsfrs}

We also utilize SFRs based on integrated-light measurements to facilitate comparisons with previous results \citep[e.g.,][]{goddard10,adamo11a}, the majority of which did not have SFHs at their disposal. We compute SFRs derived from the \ha~and FUV fluxes, which probe similar age ranges as the CMD-based measurements of 1-10 and 1-100~Myr, respectively.

The integrated-light SFRs are based on imaging from the LVL/11HUGS survey which consists of a panchromatic dataset covering GALEX UV \citep{lee11}, optical \citep{cook14a}, narrow-band \ha~\citep{kennicutt08}, and Spitzer IR \citep{dale09} wavelengths with the aim of studying both obscured and unobscured star formation in the Local Universe. All of the dwarfs studied here have imaging available from the LVL/11HUGS survey. 

We measure the \ha~and FUV fluxes from these images to derive SFRs, in an area appropriate for comparison with the star cluster populations (see \S\ref{sec:areatest} and Table~\ref{tab:tabsfr}). Photometry was performed using the \texttt{ASTROPY}\footnote{https://photutils.readthedocs.io/en/stable/} \citep{astropy13,astropy18} package \texttt{PHOTUTILS} \citep{photutils} in \texttt{PYTHON}. These fluxes are corrected for MW extinction via \cite{schlafly11} as tabulated by NED\footnote{https://ned.ipac.caltech.edu} and internal dust extinction via Spitzer 24$\mu m$ fluxes extracted inside the same apertures \citep{calzetti07,hao11}. In addition, the \ha~fluxes have been corrected for \nii~contamination using the tabulated \nii/\ha~ ratios from \cite{kennicutt08} derived via the methods described in \cite{lee09a}.

We use the SFR prescriptions of \cite{murphy11}, which are calibrated to free-free, non-extincted radio SFRs. The following assumptions were made to derive the H$\alpha$ and FUV SFRs: the Starburst99 stellar population models \citep{starburst99}, a \cite{kroupa01} IMF with a mass range of 0.1 -- 100~$M_{\odot}$, TRGB distances as measured by \cite{sabbi18}, and solar metallicity. We note that the solar metallicity assumption in the \cite{murphy11} prescriptions can result in an overestimation of H$\alpha$ and FUV SFRs by as much as a factor of 2 at low metallicities \citep{lee02,bicker05}, which is caused by the hotter temperatures, higher ionizing fluxes, and higher UV luminosities of low-metallicity stars.

\begin{table*}
{Star Formation Rates Derived inside HST FOV+D25 Ellipse}\\
\begin{tabular}{lccccccc}

\hline
\hline
Galaxy         & E(B-V) & Area      & FUV      & $\ha$      & MIPS 24$\mu m$ & SFR$_0$--FUV             & SFR$_0$--$\ha$             \\ 
Name           & (mag)  & (kpc$^2$) & (AB mag) & (AB mag)   & (AB mag)       & (M$_{\odot}$ yr$^{-1}$)  & (M$_{\odot}$ yr$^{-1}$)  \\ 
\hline

UGC7408    & 0.012 & 8.6 & 16.22 $\pm$ 0.34 & 20.30 $\pm$ 0.90 & 17.54 $\pm$ 0.42 & 5.82 $\pm$ 1.82E$-03$ & 7.89 $\pm$ 3.80E$-05$ \\
UGC5139    & 0.051 & 8.2 & 14.80 $\pm$ 0.21 & 13.71 $\pm$ 0.12 & 13.64 $\pm$ 0.07 & 6.85 $\pm$ 1.21E$-03$ & 4.76 $\pm$ 0.43E$-03$ \\
UGC4459    & 0.038 & 2.2 & 15.30 $\pm$ 0.25 & 13.67 $\pm$ 0.04 & 13.01 $\pm$ 0.05 & 5.17 $\pm$ 0.98E$-03$ & 6.39 $\pm$ 0.20E$-03$ \\
UGC4305    & 0.032 & 7.1 & 13.42 $\pm$ 0.10 & 11.70 $\pm$ 0.04 & 11.15 $\pm$ 0.02 & 2.04 $\pm$ 0.16E$-02$ & 2.39 $\pm$ 0.08E$-02$ \\
NGC5238    & 0.010 & 3.0 & 15.22 $\pm$ 0.21 & 13.81 $\pm$ 0.04 & 13.65 $\pm$ 0.07 & 6.41 $\pm$ 1.15E$-03$ & 6.69 $\pm$ 0.24E$-03$ \\
UGC5340    & 0.018 & 24.1 & 15.28 $\pm$ 0.23 & 13.68 $\pm$ 0.03 & 17.69 $\pm$ 0.45 & 4.50 $\pm$ 0.94E$-02$ & 5.71 $\pm$ 0.18E$-02$ \\
ESO486G021 & 0.034 & 5.5 & 15.45 $\pm$ 0.26 & 14.49 $\pm$ 0.05 & 13.84 $\pm$ 0.08 & 2.20 $\pm$ 0.48E$-02$ & 1.58 $\pm$ 0.07E$-02$ \\
UGC7242    & 0.019 & 2.8 & 16.14 $\pm$ 0.34 & 15.65 $\pm$ 0.27 & 14.74 $\pm$ 0.12 & 4.43 $\pm$ 1.27E$-03$ & 1.94 $\pm$ 0.40E$-03$ \\
IC559      & 0.026 & 12.8 & 16.37 $\pm$ 0.39 & 15.61 $\pm$ 0.09 & 15.13 $\pm$ 0.14 & 1.10 $\pm$ 0.37E$-02$ & 6.69 $\pm$ 0.49E$-03$ \\
NGC5477    & 0.011 & 6.5 & 14.93 $\pm$ 0.19 & 13.86 $\pm$ 0.04 & 13.40 $\pm$ 0.06 & 1.90 $\pm$ 0.30E$-02$ & 1.51 $\pm$ 0.05E$-02$ \\
NGC4248    & 0.020 & 8.1 & 16.71 $\pm$ 0.44 & 14.97 $\pm$ 0.06 & 12.77 $\pm$ 0.05 & 6.72 $\pm$ 1.43E$-03$ & 8.00 $\pm$ 0.38E$-03$ \\
UGC685     & 0.057 & 1.3 & 16.04 $\pm$ 0.37 & 14.52 $\pm$ 0.06 & 14.41 $\pm$ 0.10 & 2.95 $\pm$ 0.91E$-03$ & 3.40 $\pm$ 0.16E$-03$ \\
NGC1705    & 0.008 & 4.8 & 13.38 $\pm$ 0.09 & 12.15 $\pm$ 0.04 & 12.18 $\pm$ 0.04 & 4.71 $\pm$ 0.37E$-02$ & 4.44 $\pm$ 0.15E$-02$ \\
UGCA281    & 0.015 & 0.9 & 15.23 $\pm$ 0.22 & 12.85 $\pm$ 0.02 & 12.15 $\pm$ 0.04 & 1.12 $\pm$ 0.16E$-02$ & 2.35 $\pm$ 0.05E$-02$ \\
IC4247     & 0.066 & 1.1 & 15.92 $\pm$ 0.36 & 16.03 $\pm$ 0.21 & 15.40 $\pm$ 0.16 & 4.22 $\pm$ 1.36E$-03$ & 1.18 $\pm$ 0.18E$-03$ \\
UGC695     & 0.028 & 4.1 & 16.63 $\pm$ 0.44 & 15.45 $\pm$ 0.22 & 14.92 $\pm$ 0.13 & 5.50 $\pm$ 1.99E$-03$ & 4.16 $\pm$ 0.73E$-03$ \\
UGC1249    & 0.079 & 24.9 & 13.67 $\pm$ 0.14 & 13.14 $\pm$ 0.03 & 12.09 $\pm$ 0.03 & 5.58 $\pm$ 0.63E$-02$ & 2.89 $\pm$ 0.06E$-02$ \\
NGC3274    & 0.024 & 14.5 & 14.45 $\pm$ 0.16 & 13.24 $\pm$ 0.03 & 12.01 $\pm$ 0.03 & 7.40 $\pm$ 0.88E$-02$ & 6.61 $\pm$ 0.15E$-02$ \\
NGC5253    & 0.056 & 4.9 & 12.29 $\pm$ 0.07 & 9.95 $\pm$ 0.01 & 6.58 $\pm$ 0.00 & 2.77 $\pm$ 0.03E$-01$ & 3.35 $\pm$ 0.01E$-01$ \\
NGC4485    & 0.022 & 19.1 & 13.61 $\pm$ 0.11 & 12.00 $\pm$ 0.04 & 10.76 $\pm$ 0.02 & 1.35 $\pm$ 0.10E$-01$ & 1.43 $\pm$ 0.05E$-01$ \\
NGC3738    & 0.010 & 7.1 & 13.78 $\pm$ 0.11 & 12.54 $\pm$ 0.06 & 11.24 $\pm$ 0.02 & 3.62 $\pm$ 0.29E$-02$ & 2.94 $\pm$ 0.13E$-02$ \\
NGC4449    & 0.019 & 10.4 & 11.04 $\pm$ 0.03 & 9.51 $\pm$ 0.01 & 7.81 $\pm$ 0.00 & 3.31 $\pm$ 0.07E$-01$ & 3.28 $\pm$ 0.03E$-01$ \\
NGC4656    & 0.013 & 37.9 & 12.62 $\pm$ 0.07 & 11.56 $\pm$ 0.01 & 10.14 $\pm$ 0.01 & 2.51 $\pm$ 0.12E$-01$ & 2.01 $\pm$ 0.02E$-01$ \\
\hline

\end{tabular}
\caption{The integrated-light photometry and dust-corrected star formation rates for the LEGUS dwarf galaxy sample. The areas listed are those computed inside the intersection of the D25 ellipse (see \S\ref{sec:areatest}) and the HST FOV (\aper). SFRs are derived via the prescriptions of \citet{murphy11} and are corrected for MW extinction and internal dust extinction via Spitzer 24$\mu m$ fluxes extracted inside the same apertures \citep{calzetti07,hao11}. In addition, the \ha~values have been corrected for \nii~contamination using the data from \citet{kennicutt08}. The galaxies are sorted by \sfrsig~using the 10-100~Myr SFHs and the \aper~areas (see \S\ref{sec:areatest}).}
\label{tab:tabsfr}

\end{table*}

\subsubsection{SFR Comparisons} \label{sec:sfrcomp}

\begin{figure}
  \begin{center}
  \includegraphics[scale=0.52]{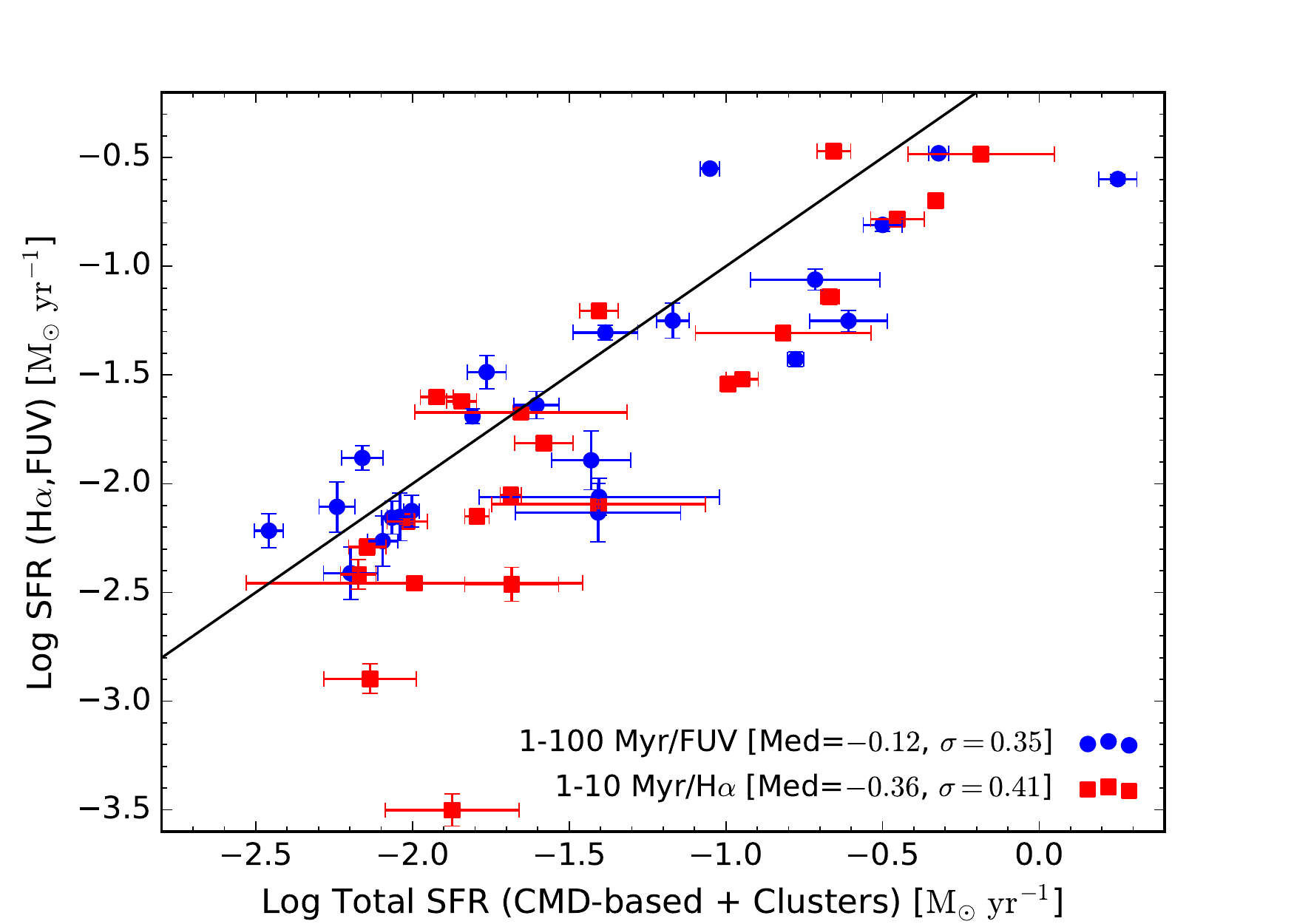}
  \caption{Comparison of SFRs from \ha/FUV measurements and those derived from counts of young stars via CMD-fitting, stellar clusters, and associations (“total stellar population SFRs”). All SFRs are measured within the HST FOV area, assume the same distances, and have been corrected for internal dust attenuation. The solid black line is a 1-to-1 correlation line. The \ha~and FUV SFRs are compared to the mass of stars and clusters integrated over 1-10 and 1-100~Myr, respectively. The FUV SFRs show overall agreement with the corresponding total stellar population values, but the \ha~SFRs tend to be underestimated compared to the 1-10~Myr total stellar population values. }
   \label{fig:sfrcomp}
   \end{center}
\end{figure}  

Figure~\ref{fig:sfrcomp} shows the comparison of SFRs derived from \ha~and FUV fluxes (vertical axis) and those based on the resolved stellar populations (horizontal axis), where we assume that the \ha~SFRs have a timescale of a few 10$^6$ year and the FUV SFRs probe timescales on order of 10$^8$ years \citep{kennicutt98,salim07,lee09b,pflam09}. We note that this SFR comparison is reasonably appropriate given the similar assumptions made for the two methods: the same TRGB distances, a Kroupa IMF, and that the \ha~and FUV fluxes are measured within the same HST footprint from which the stellar populations are identified and their properties measured. 

We find a similar scatter of 0.3--0.4~dex for both age ranges (i.e., a factor of 2), and that both \ha~and FUV SFRs are on average underestimated compared to the total stellar population values. There is only a slight underestimate of $\approx$0.1~dex between the FUV and the corresponding 1-100~Myr total stellar population SFRs (blue points) which agrees with previous comparisons \citep{mcquinn15,chandar17,cignoni19}. However, the \ha~SFRs are offset by a factor of 2 (especially at lower SFRs) compared to the 1-10~Myr total stellar population SFRs (red points). 


The larger discrepancy between the \ha\ and UV SFRs at low SFRs has been discussed extensively in prior work, with a range of explanations proposed \citep[e.g. stochastic sampling of the stellar IMF; IMF deficient in high mass stars in low density environments; time variable star formation histories; Integrated Galactic IMF theory;][]{lee09b,meurer09,pflam09,weisz12}. SFRs measured from resolved stellar populations (stellar CMDs $+$ star cluster populations) appear to be more robust to these issues, as supported by Figure 1. We note that we can confirm larger discrepancies in the \ha-based SFRs at lower SFRs compared to the FUV-derived values (plot not shown for brevity). A further complication stems from a potential overestimation of the \ha-based SFRs due to metallicity effects by up to a factor of 2 for the lowest metallicity galaxies which also tend to have low SFRs, and would further exacerbate the discrepancy. However, we note that most of our galaxies have metallicities of SMC or greater (two, three, and fourteen have metallicities of solar, LMC, and SMC, respectively), and only 4 have lower metallicities (i.e., Z=0.001 for UGC 4459, UGC 5139, UGC 5340, and UGCA 281).


In summary, we expect the \ha-based SFRs to be low  at lower SFRs compared to both the FUV-based and total stellar population SFRs, and that both the \ha~and FUV SFRs are likely to be overestimated to some degree due to metallicity effects in lower metallicity galaxies. It is important to take these caveats into account when examining the results presented in \S\ref{sec:results}. However, we note that the main results of this study are unaffected (i.e., are similar) when using  \ha~and FUV SFRs and those from the total stellar population SFRs. 

An examination of different SFR methods is important in and of itself, but in the context of cluster-host relationships we ultimately need to address the scatter in the SFR densities (\sfrsig) which involves both the SFRs and the areas within which they are calculated. In  \S\ref{sec:areatest}, we further examine the effects of area where there can be significant \sfrsig~offsets (up to 0.7 dex) depending on the area used, and choose a fiducial area to best match host galaxy and star cluster properties.

\section{Star Cluster Properties} \label{sec:clustcat}
In this section we briefly describe the LEGUS star cluster catalogs which include morphological classification, photometry, and SED fitting. For more details, see \cite{adamo17} and C19, which provide a full description of the catalog construction. We note that star cluster catalogs for 5 additional galaxies have been completed and added to this analysis compared to C19 which completes the LEGUS dwarf and irregular galaxy sample; these galaxies include: ESO486-G021, NGC 3274, NGC 4485, UGC 5340, and UGC 7242.

\subsection{Cluster Catalogs}

Initial cluster candidates were culled from an SExtractor catalog.  Candidates were retained if they showed an extended light profile as measured via a concentration index \citep[CI; defined as the magnitude difference at 1 and 3 pixels;][]{chandar10a}. These candidates were then visually inspected by three or more team members and given numerical classifications: class 1 objects are extended sources with spherical symmetry; class 2 object are extended sources, but have some degree of asymmetry in their radial profiles; class 3 objects are those with multiple peaks in their radial profiles; class 4 objects are those considered to be contaminants (e.g., obvious stars, background galaxies, random overdensities of nebular emission, etc.). These morphology classifications potentially provide insight into the evolutionary status of the clusters.  Class 1 and 2 objects may be gravitationally bound star clusters assuming their ages are greater than the crossing time needed to disperse the stellar population \citep[$\approx$10~Myr;][]{gieleszwart11,krumholz19}. Class 3 objects show multiple stellar peaks and are likely compact stellar associations, which may be in the process of being dissolved \citep{grasha15}.

In this analysis we use only the likely bound clusters when measuring cluster properties as one of the main parameters of interest in this study, $\Gamma$. Overall, the total number of clusters (Class 1 and 2) at all ages in this sample of dwarf galaxies is 1371.  It should be noted that 4 of the 23 galaxies  contain $\approx$70\% of the clusters (N=321 in NGC 4449, N=298 in NGC 4485, N=184 in NGC 4656, and N=141 in NGC 3738). The sample as a whole shows a median of 21 clusters per galaxy. 

The total fluxes tabulated for star clusters in the LEGUS cluster catalogues are computed from aperture photometry in two ways: with an average aperture correction derived from training clusters, and a CI-based aperture correction which takes into account the extent of each cluster's light profile (C19). 
The ensemble cluster colors, luminosity and mass functions in the dwarf galaxies are not affected by the type of aperture correction used (C19).  However, the total masses for individual clusters can be discrepant by as much as several tenths of a dex, which may have a significant effect on $\Gamma$ measurements in galaxies with only a few clusters. In this analysis we utilize the CI-based aperture correction to mitigate these effects. However, we note that we find similar results when using the average aperture corrected photometry.

The cluster ages and masses are determined via SED fitting to single-aged stellar population models. The methods are detailed fully in \citet{adamo17}, but we provide a brief overview here. We utilize the Yggdrasil \citep{zackrisson11} SSP models with the assumption that the IMF is fully sampled, Padova-AGB tracks \citep{tang14}, a Kroupa IMF that ranges from 0.1 to 100~$\rm{M}_{\odot}$, a starburst attenuation curve with differential reddening \citep{calzetti00}, an escape fraction of 0.5, and the measured gas phase metallicity of each galaxy. These models are input into Cloudy \citep{cloudy} to produce fluxes from nebular emission lines and continuum. 

\subsection{Updates to SED Fitting}

With respect to the analysis reported in \cite{cook19a}, an error in the produced SED models has surfaced, which has affected the recovered physical parameters of about 15 \% of the clusters. When producing the final SSP spectral models (before convolution with filter transmissions) the contribution of nebular emission spectra produced by Cloudy \citep{cloudy} has only partially been accounted for. The resulting integrated magnitudes of passbands containing strong emission lines (namely, F555W and F606W) have been underestimated in the age steps corresponding to the highest nebular emission contribution phase. Because of the nature of the error, the problem has affected the resulting magnitudes of our SSP models at the youngest age steps ($<$ 5 Myr) and only at metallicity of Z=0.008 and lower, therefore relevant for this study. 

New SED fits have been repeated for all the LEGUS cluster catalogues included in this work and have been updated on the LEGUS website\footnote{https://legus.stsci.edu/}. The new fits have produced older cluster ages for about 15\% of the clusters in this sample. The main change regards clusters that have been initially assigned as best-fitted age 1~Myr. Roughly 28\% of these clusters have now moved to 3~Myr. Another significant change is noticed in a smaller fraction ($\approx14$\%) of clusters with initial best age of 5 Myr, which have moved into the 10~Myr range, after the correct SSP models have been used in the fit. Neither of these changes affect our results since the affected clusters remain in the same age ranges used in our study (1-10~Myr). In addition, the recovered masses and extinctions do not change significantly and remain within the mass uncertainties \citep[about 0.2 dex, see][]{adamo17}. The results reported in \cite{cook19a} remain unchanged when using the updated cluster catalogues.

\subsection{Completeness Limits} \label{sec:clustlimits}

\begin{figure}
  \begin{center}
  \includegraphics[scale=0.52]{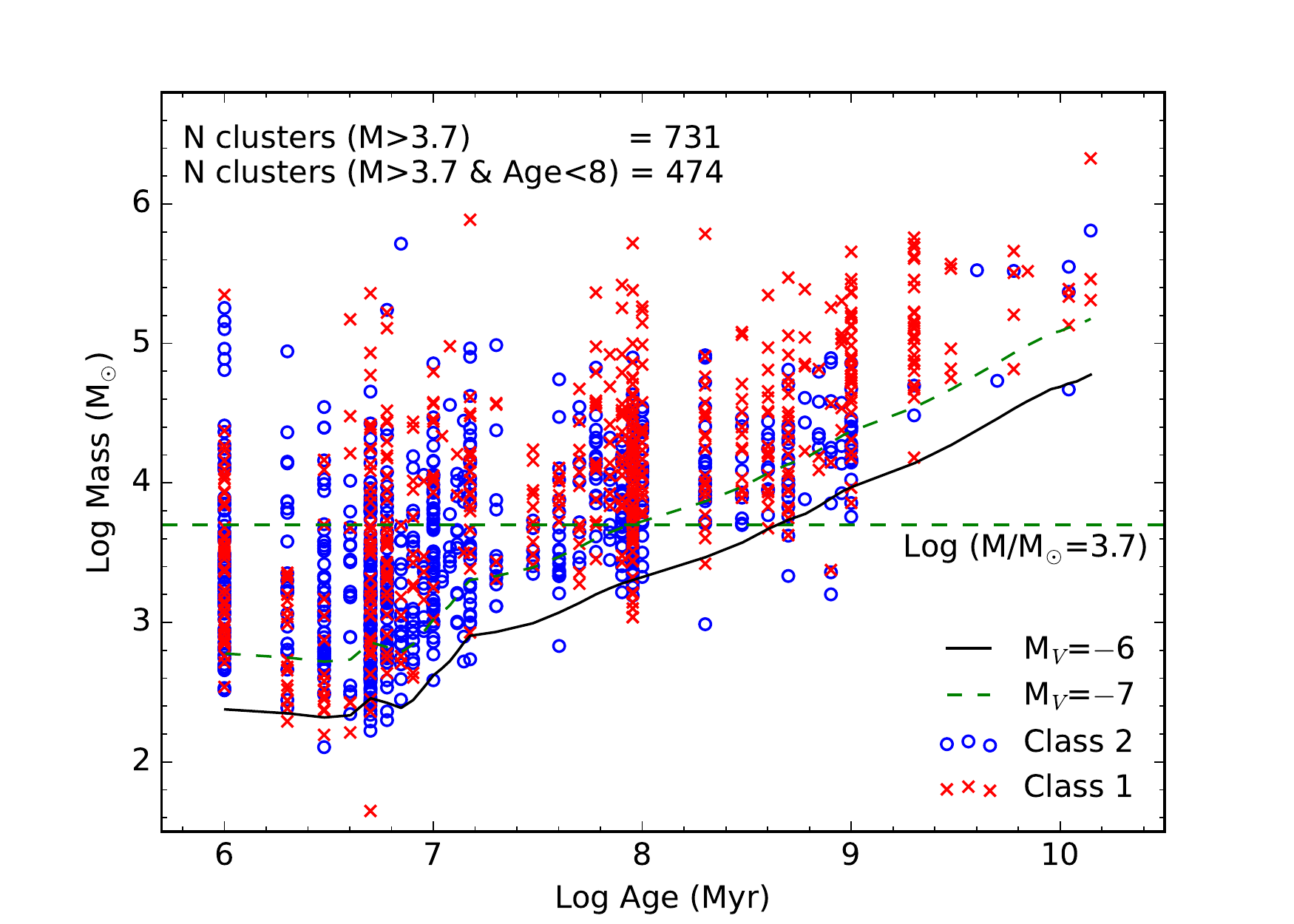}
  \caption{The star cluster mass as a function of age for class 1 and 2 clusters (N=1371) for all galaxies in our sample. The solid and dashed curves represent the stellar masses which correspond to absolute $V-$band magnitudes of $-6$ and $-7$~mag, respectively, for single-aged stellar populations. In this study we adopt a completeness limit of --7~mag determined in C19. The horizontal line shows the mass limit of Log($M$/M$_{\odot})=$3.7 which corresponds to where the --7~mag curve meets the maximum cluster age examined in this study ($<$100~Myr).}
   \label{fig:agemass}
   \end{center}
\end{figure}

Figure~\ref{fig:agemass} show the age-mass diagram for all class 1 and 2 clusters in our dwarf galaxy sample. Using simple stellar population models, we have plotted --6 and --7~mag curves to assess completeness. Visual inspection shows that the majority of the clusters are brighter than M$_V=-6$~mag detection threshold imposed on all LEGUS galaxies, but the completeness limit is brighter than this. The C19 study showed that the clusters in this sub-sample are largely complete at an absolute $V-$band magnitude of $-$7~mag, which coincides with the turnover seen in the LFs for individual galaxies (see Appendix~\ref{sec:append_lfmfdndt} for examples) and in the composite luminosity function using the clusters from all galaxies. 

Applying this absolute $V-$band mag limit of --7~mag to the age-mass diagram, we can then define the corresponding mass limit to produce a mass-limited cluster sample. However, this needs to be applied to a given age range of clusters. Choosing an age limit of $<$100~Myr for consistency with previous cluster-host relationships \citep{larsen02,bastian08,goddard10,adamo11c,cook12}, we then determine a mass limit of Log($M$/M$_{\odot})=$3.7 which is defined by the intersection of the --7~mag curve at 100~Myr. The mass limit is denoted as the dashed, horizontal line in Figure~\ref{fig:agemass}.

\section{Galaxy Spatial Coverage} \label{sec:areatest}


The choice of area over which to study cluster-host galaxy relationships can have a large effect on their normalization and scatter. Here we describe the process used to determine the area over which the SFRs are measured and that spatially coincides with the cluster populations captured by the LEGUS HST observations. 


\begin{figure*}
  \begin{center}
  \includegraphics[scale=0.42]{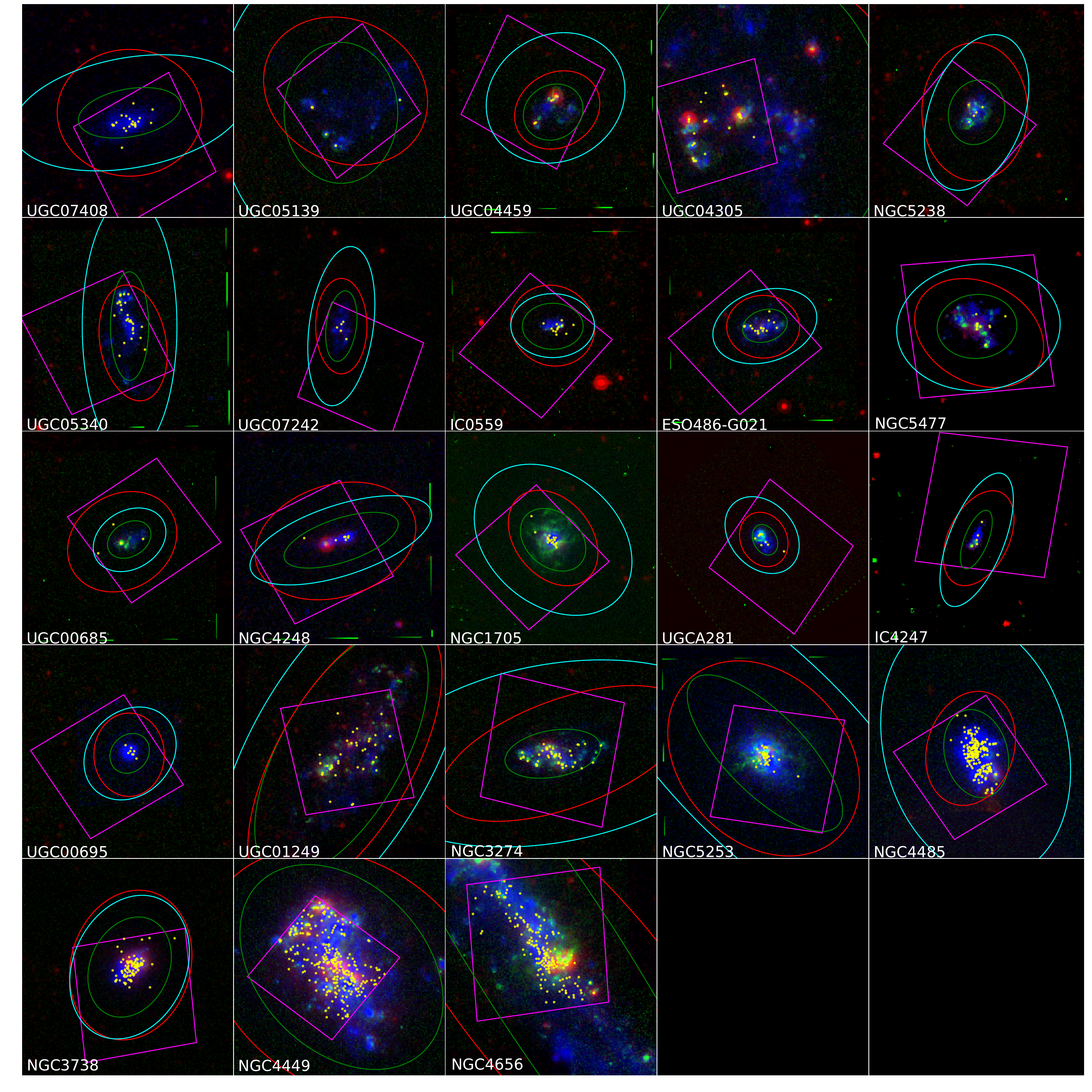}
  \caption{Color images for the dwarf galaxies used in this study, where GALEX FUV is coded blue, ground-based \ha~is coded green, and Spitzer $24\mu m$ is coded red. Three galaxy-wide apertures are shown which are derived from the extent of the GALEX FUV fluxes shown as cyan ellipses, B-band fluxes at 25th mag/arcsec$^2$ (D25) shown as green ellipses, and Spitzer IRAC 1 fluxes shown as red ellipses. The UV, D25, and IR ellipses roughly correspond to the extended low-surface brightness emission, optical emission, and the stellar mass, respectively. The images are centered on the D25 aperture and the coverage extends roughly to encompass the HST FOV as indicated by the magneta squares. The small yellow symbols represent the class 1 and 2 clusters with ages less than 100~Myr. The galaxies are sorted by \sfrsig~using the 10-100~Myr SFHs and the D25$\cap$HST FOV areas (see \S\ref{sec:areatest}). }
   \label{fig:colorPics}
   \end{center}
\end{figure*} 

Figure~\ref{fig:colorPics} shows FUV-\ha-IR (blue-green-red) color images of the galaxies in our sample along with 3 ellipses from the literature that have been used to define the extent of nearby galaxies: 1) the RC3 catalog isophotal aperture defined at the $B-$band 25 mag/arcsec$^2$ surface brightness \citep[hereafter D25 apertures; ][]{rc3}, 2) the apertures defined by the asymptotic extent of the GALEX FUV emission \cite[hereafter UV apertures;][]{lee11}, and 3) the apertures defined by capturing all of the  galaxy emission visible for all infrared images from 3.6--160$\mu m$ \citep[hereafter IR apertures;][]{dale09}. 

\begin{figure}
  \begin{center}
  \includegraphics[scale=0.45]{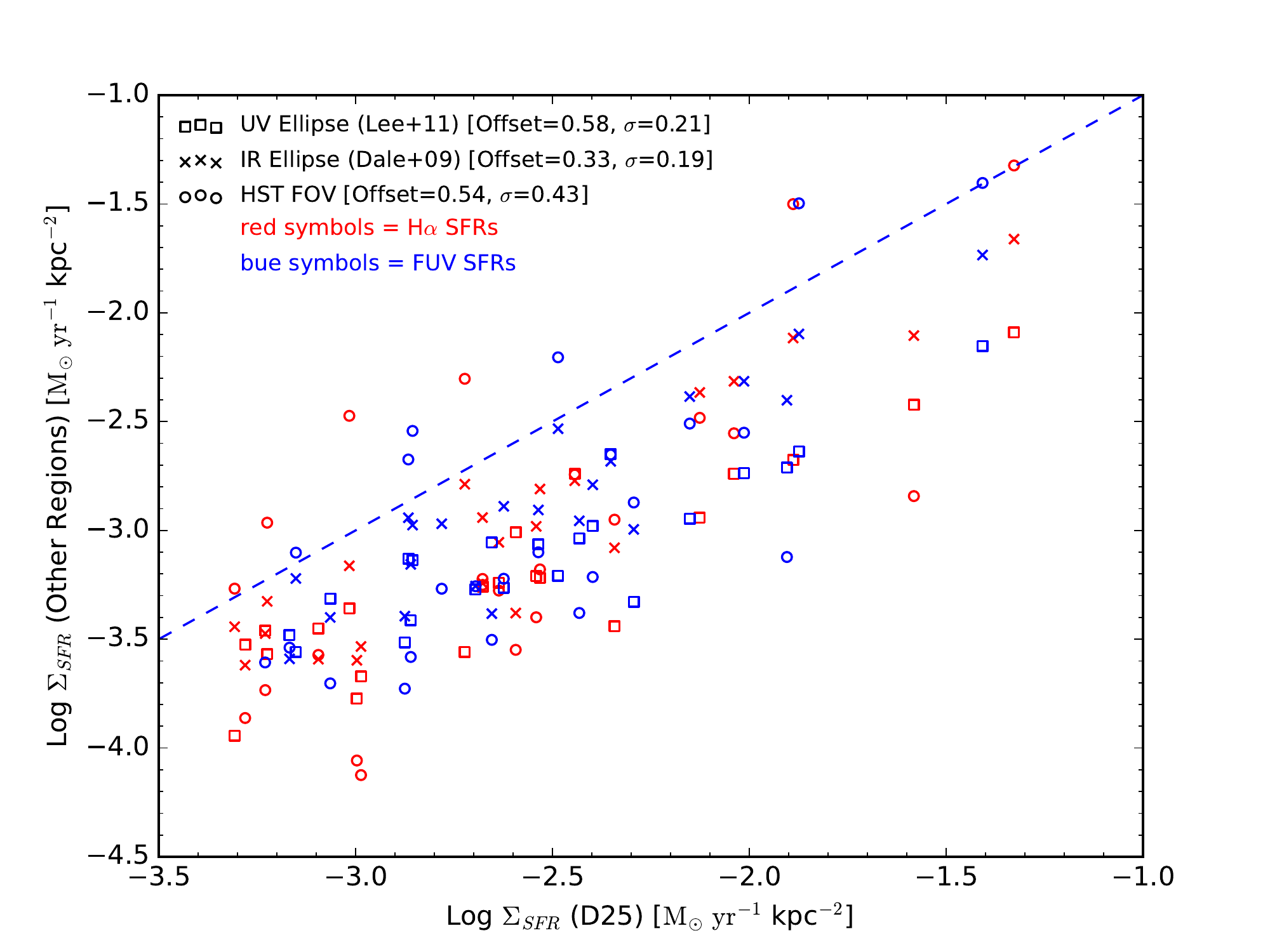}
  \caption{Comparison of \sfrsig~using different galaxy area definitions represented as different symbols, where the x-axis is the more commonly used D25 aperture. The SFRs are derived from \ha~and FUV fluxes (corrected for dust via Spitzer 24$\mu m$ fluxes) measured inside the specified aperture which are differentiated as red and blue symbols, respectively. The median offsets can be significant with values reaching 0.6~dex. }
   \label{fig:sigmacomp}
   \end{center}
\end{figure}

In short, the D25, UV, and IR apertures effectively encompass the optical emission of the galaxy, the low surface brightness outskirts, and the majority of the total stellar mass, respectively. Visual inspection of these apertures in Figure~\ref{fig:colorPics} show that the UV apertures are typically larger than the IR apertures which are typically lager than the D25 apertures. The \sfrsig~values derived from these various apertures will show differences when compared to each other. Figure~\ref{fig:sigmacomp} compares the \sfrsig~values computed within the UV apertures, IR apertures, and HST FOV to the more commonly used D25 aperture. The UV, IR, and HST FOV all show \sfrsig~values that are offset to smaller values (0.3--0.6 dex) which demonstrates that the choice of galaxy aperture can have a significant effect on \sfrsig~and consequently the normalization of cluster-host relationships. We also note that the HST FOV aperture shows significantly larger scatter ($\sigma\approx0.4$) compared to other apertures which is likely due to varying HST coverage (i.e., only covers some of the galaxy or covers more than the entire galaxy).

\begin{figure}
  \begin{center}
  \includegraphics[scale=0.39]{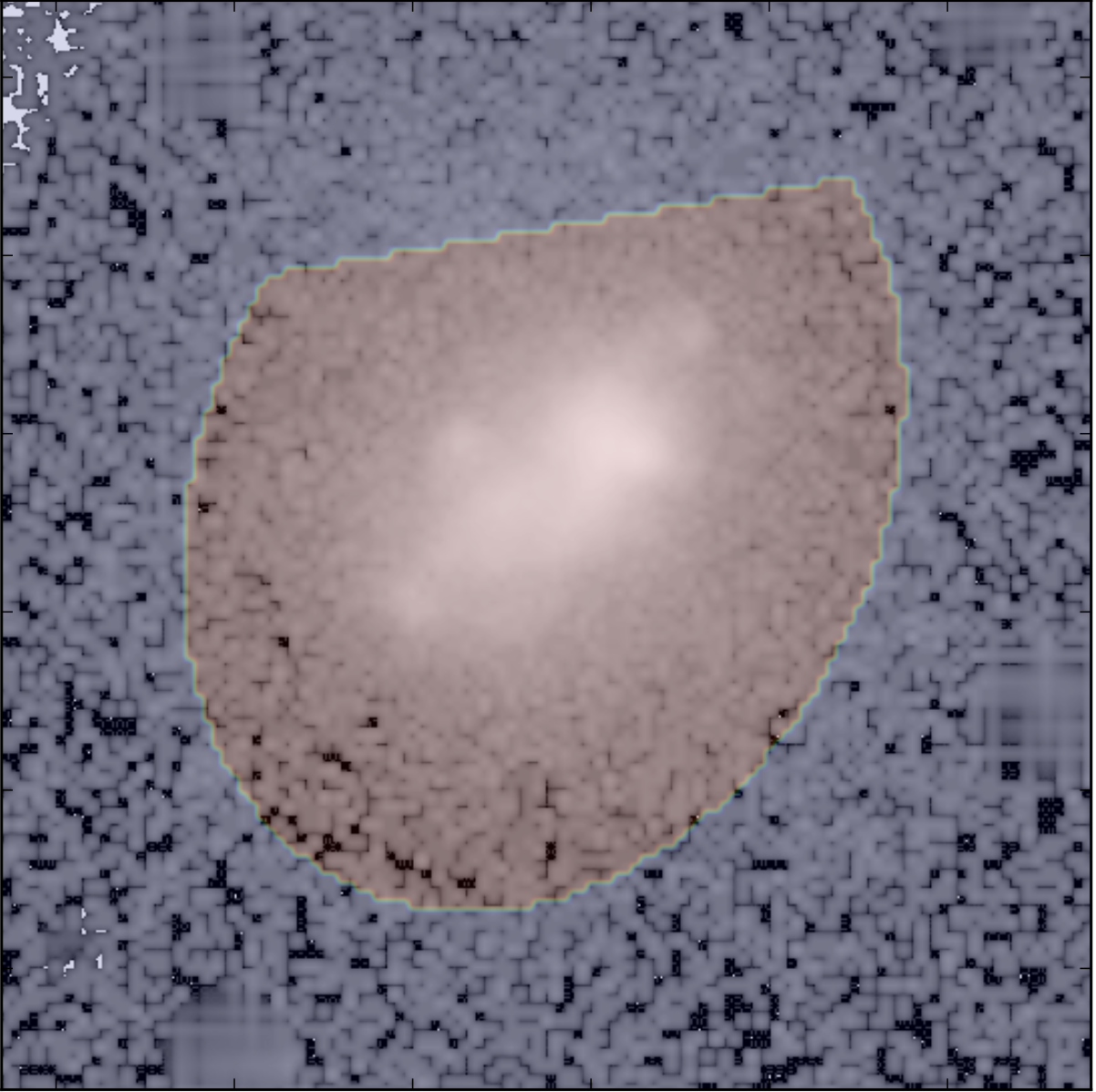}
  \caption{The gray-scale GALEX FUV image of NGC 3738 where the shaded region is the final area aperture which is a convolution between the HST FOV and the D25 ellipse (\aper). The top portion of the D25 ellipse is cut off due to the HST FOV (see the bottom-left panel of Figure~\ref{fig:colorPics}).}
   \label{fig:convoEx}
   \end{center}
\end{figure}

An appropriate aperture over which to compute global properties in the study of cluster-host galaxy relationships for our sample can be determined by examination of the cluster population. The clusters were identified via the HST imaging, thus they will all be inside HST FOV.  Furthermore, we find that 98.7\% (1353 out of 1371) of the clusters lie within the D25 aperture. We conclude that the appropriate aperture is the intersection ($\cap$) of the D25 ellipses and the HST FOV (hereafter \aper), which is the aperture adopted for our study of cluster-host galaxy relationships. Figure~\ref{fig:convoEx} shows an example for NGC3738 where the HST FOV intersects with the top portion of the D25 ellipse. Table~\ref{tab:tabsfr} presents the MW- and internal dust-corrected SFRs and areas of the sample inside the \aper~regions.

\begin{figure}
  \begin{center}
  \includegraphics[scale=0.45]{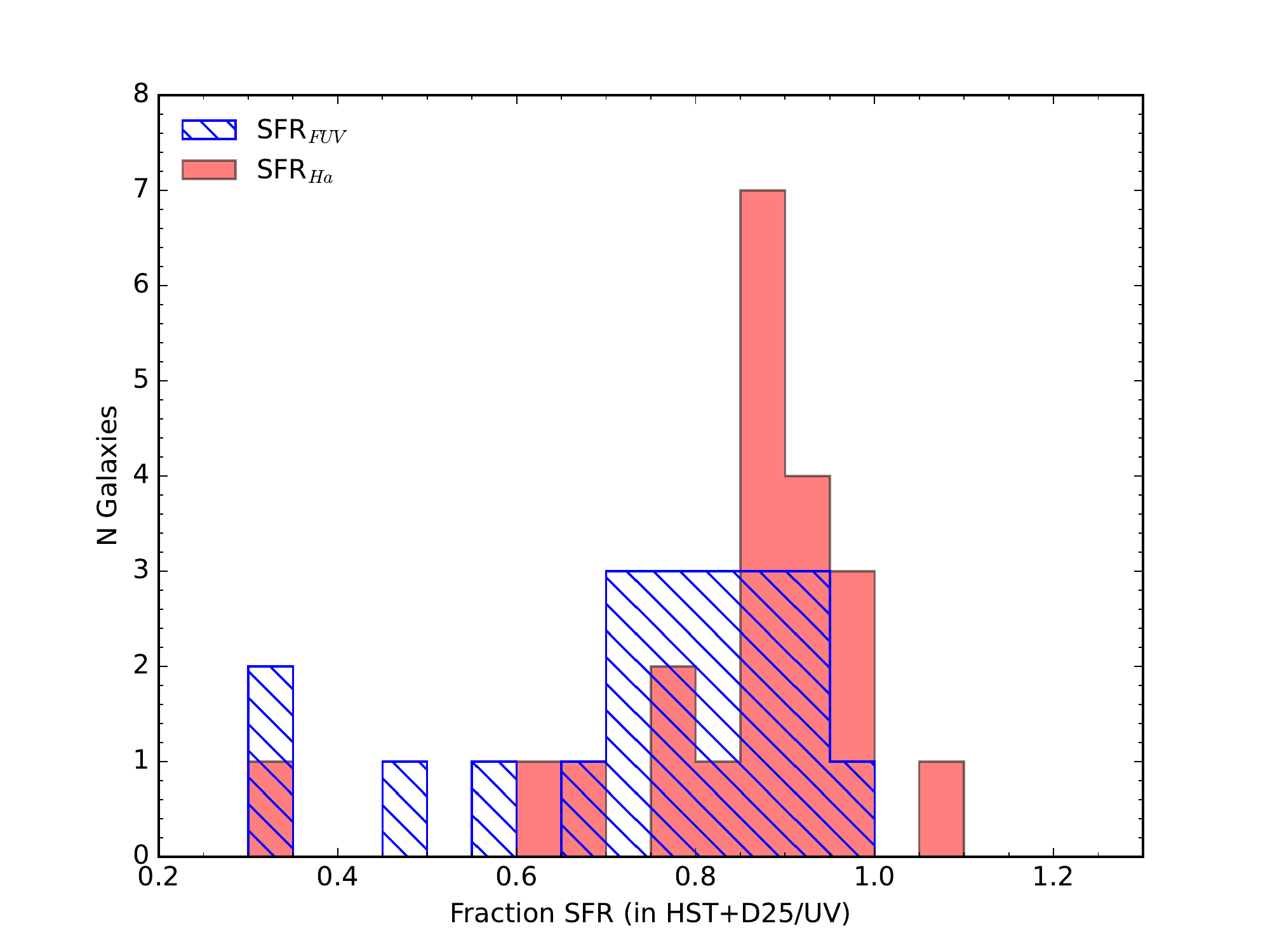}
  \caption{Fraction of the dust-corrected SFRs derived inside the \aper~and UV regions where the UV regions are assumed to approximate each galaxy's total SFR. We find that the \ha~SFRs are more concentrated while the UV SFRs are more dispersed. UGC~695 shows an \ha-based SFR ratio greater than one, but this is likely due to relatively higher measurement errors for this low SFR galaxy ($\approx10^{-3}~\sfrunit$) where we find SFRs for both apertures that agree within the uncertainties. }
   \label{fig:fraction}
   \end{center}
\end{figure}

It is interesting to compute the fraction of the total star formation activity captured by our adopted \aper~regions, and examine the range of coverage fractions for our LEGUS dwarf and irregular galaxy sub sample. Figure~\ref{fig:fraction} shows the SFR ratios derived from our adopted \aper~region-to-UV areas, where we assume that the SFR measured inside the UV aperture is a proxy for the total SFR given their larger sizes (see Figure~\ref{fig:colorPics} for aperture examples). The peak near 0.9 of the \ha~SFR-to-total fraction in our chosen measurement aperture (\aper) suggests that very little recent star formation (1--10~Myr) occurs in the outer regions. Conversely, the FUV-to-total SFR fractions (effectively a comparison of the FUV flux measured within two different apertures) are not strongly peaked suggesting that older (1--100~Myr) star formation is more evenly distributed over the entire galaxy. This conclusion is in agreement with previous studies that have shown more extended UV emission in nearby galaxies compared to their optical extent  \citep[e.g.,][]{thilker07}, which could be due to $in situ$ star formation in the outskirts or the smoothing of clustered star formation over timescales that roughly correspond to those of orbital motions.

There is one galaxy (UGC 695) that shows an \ha-based SFR ratio slightly higher than 1 which seemingly contradicts our assumption that the SFR measured inside the UV area is an approximation for the total SFR. However, this discrepancy is likely due to the relatively low \ha~SFR and subsequent higher measurement errors. We find that both the D25 and UV apertures cover the extent of the \ha~flux~and that the resulting SFRs are in agreement within the errors: SFR$=4.1~\rm{and}~4.3 \pm 0.7~\times 10^{-3}~\sfrunit$ for UV and D25 apertures, respectively. There are also several galaxies that show small SFR ratios (i.e., less 50\%). This low ratio is due to HST coverage that does not cover the extent of the galaxy's flux in \ha~and/or UV (see NGC 4656, UGC 1249, UGC 4305, UGC 5139 in Figure~\ref{fig:colorPics}). 

While we have chosen a specific aperture here to facilitate comparisons to previous studies, we note that there is still an open question of what physical scales are relevant for cluster-host relationships. Specifically, do these relationships hold at smaller physical scales and subsequently what physical mechanisms drive them? This analysis is left for a second paper in this series and is currently underway.

\section{Results}  \label{sec:results}
In this section we present several star cluster-host galaxy relationships previously established in the literature and concentrate on the fraction of stars in clusters ($\Gamma$)-\sfrsig~relationship. We find that the cluster populations of dwarf galaxies follow established size-of-sample relationships while other cluster properties show no statistically significant trend with global galaxy properties.

\subsection{M$_{V}^{\rm Brightest}$ -- SFR Relationship} \label{sec:MvSFR}

The relationship between the brightest cluster and the host-galaxy's SFR has been studied by numerous authors in a wide variety of galaxy environments spanning dwarfs to spirals to starbursts \citep{Larsen10,bastian16,krumholz19}. Related to the number of clusters versus SFR, this trend has been attributed to a size-of-sample effect where higher SFR galaxies produce greater numbers of clusters, but there has also been extensive discussion on whether random sampling alone is consistent with the observations or whether the data implicate other physical processes that for example limit the maximum mass of clusters \citep{whitmore00,larsen02,weidner04,bastian08,adamo11c,cook12,Randriamanakoto13,whitmore14a}.

\begin{table*}

{Star Cluster Statistics in LEGUS Dwarf Galaxies}\\
\begin{tabular}{lrccrccrcc}

\hline
\hline
Galaxy     & \multicolumn{3}{c|}{1--10 Myr}                  &  \multicolumn{3}{c|}{1--100 Myr}                 & \multicolumn{3}{c|}{10--100 Myr}                 \\
Name       & N    & M$_{V}^{bright}$ & CFR                   & N    & M$_{V}^{bright}$  & CFR                   & N    & M$_{V}^{bright}$ & CFR                    \\
           & (\#) & (mag)            & ($M_{\odot} yr^{-1}$) & (\#) & (mag)             & ($M_{\odot} yr^{-1}$) & (\#) & (mag)            & ($M_{\odot} yr^{-1}$)   \\

\hline

UGC7408 & 5 & -8.05 $\pm$ 0.27 & 7.3 $\pm$ 4.6E$-03$ & 22 & -8.77 $\pm$ 0.10 & 5.1 $\pm$ 0.6E$-03$ & 17 & -6.36 $\pm$ 0.09 & 5.0 $\pm$ 0.5E$-03$ \\
UGC5139 & 3 & -6.94 $\pm$ 0.10 & \ldots & 3 & -6.94 $\pm$ 0.10 & \ldots & 0 & \ldots & \ldots \\
UGC4459 & 3 & -7.16 $\pm$ 0.10 & \ldots & 3 & -7.16 $\pm$ 0.10 & \ldots & 0 & \ldots & \ldots \\
UGC4305 & 7 & -8.63 $\pm$ 0.08 & \ldots & 8 & -8.63 $\pm$ 0.08 & \ldots & 1 & -6.02 $\pm$ 0.31 & \ldots \\
NGC5238 & 2 & -10.69 $\pm$ 0.09 & 1.0 $\pm$ 0.1E$-02$ & 3 & -10.69 $\pm$ 0.09 & 1.1 $\pm$ 0.1E$-03$ & 1 & -7.27 $\pm$ 0.25 & 3.7 $\pm$ 0.6E$-04$ \\
UGC5340 & 14 & -9.23 $\pm$ 0.10 & 1.5 $\pm$ 0.2E$-02$ & 29 & -11.15 $\pm$ 0.04 & 9.6 $\pm$ 1.0E$-03$ & 15 & -11.15 $\pm$ 0.04 & 9.7 $\pm$ 1.0E$-03$ \\
ESO486G021 & 7 & -6.94 $\pm$ 0.08 & 8.5 $\pm$ 15.1E$-03$ & 11 & -8.15 $\pm$ 0.24 & 1.5 $\pm$ 1.4E$-03$ & 4 & -6.07 $\pm$ 0.08 & 1.3 $\pm$ 0.1E$-03$ \\
UGC7242 & 2 & -6.91 $\pm$ 0.08 & \ldots & 5 & -6.91 $\pm$ 0.08 & 3.5 $\pm$ 0.0E$-04$ & 3 & -6.37 $\pm$ 0.03 & 3.8 $\pm$ 0.0E$-04$ \\
IC559 & 4 & -8.24 $\pm$ 0.07 & 1.1 $\pm$ 0.1E$-02$ & 11 & -8.24 $\pm$ 0.07 & 2.8 $\pm$ 0.1E$-03$ & 7 & -7.29 $\pm$ 0.07 & 2.2 $\pm$ 0.1E$-03$ \\
NGC5477 & 5 & -7.24 $\pm$ 0.11 & 3.8 $\pm$ 3.7E$-03$ & 6 & -7.24 $\pm$ 0.11 & 3.5 $\pm$ 3.3E$-04$ & 1 & -6.38 $\pm$ 0.18 & \ldots \\
NGC4248 & 3 & -7.37 $\pm$ 0.07 & 7.8 $\pm$ 1.7E$-03$ & 6 & -9.08 $\pm$ 0.04 & 4.3 $\pm$ 2.2E$-03$ & 3 & -6.90 $\pm$ 0.04 & 4.2 $\pm$ 2.5E$-03$ \\
UGC685 & 2 & -7.99 $\pm$ 0.31 & 8.0 $\pm$ 11.5E$-03$ & 2 & -7.99 $\pm$ 0.31 & 7.3 $\pm$ 10.5E$-04$ & 0 & -4.81 $\pm$ 0.09 & \ldots \\
NGC1705 & 12 & -13.16 $\pm$ 0.17 & 1.0 $\pm$ 1.2E$-01$ & 20 & -13.16 $\pm$ 0.17 & 1.2 $\pm$ 1.1E$-02$ & 8 & -7.27 $\pm$ 0.25 & 3.5 $\pm$ 0.6E$-03$ \\
UGCA281 & 4 & -9.33 $\pm$ 0.10 & 3.7 $\pm$ 0.0E$-03$ & 4 & -9.33 $\pm$ 0.10 & 3.3 $\pm$ 0.0E$-04$ & 0 & -4.60 $\pm$ 0.09 & \ldots \\
IC4247 & 2 & -7.02 $\pm$ 0.27 & \ldots & 5 & -8.10 $\pm$ 0.28 & 9.1 $\pm$ 4.8E$-04$ & 3 & -8.10 $\pm$ 0.28 & 1.0 $\pm$ 0.5E$-03$ \\
UGC695 & 3 & -6.99 $\pm$ 0.24 & 5.2 $\pm$ 5.8E$-03$ & 5 & -7.93 $\pm$ 0.03 & 1.1 $\pm$ 0.6E$-03$ & 2 & -6.13 $\pm$ 0.08 & 7.6 $\pm$ 0.4E$-04$ \\
UGC1249 & 13 & -9.16 $\pm$ 0.10 & 7.4 $\pm$ 2.8E$-03$ & 28 & -9.16 $\pm$ 0.10 & 4.1 $\pm$ 0.4E$-03$ & 15 & -7.34 $\pm$ 0.10 & 4.0 $\pm$ 0.3E$-03$ \\
NGC3274 & 12 & -10.25 $\pm$ 0.07 & 2.5 $\pm$ 0.7E$-02$ & 20 & -10.25 $\pm$ 0.07 & 4.3 $\pm$ 0.8E$-03$ & 8 & -6.02 $\pm$ 0.09 & 2.8 $\pm$ 0.4E$-03$ \\
NGC5253 & 17 & -10.80 $\pm$ 0.18 & 5.3 $\pm$ 2.0E$-02$ & 29 & -10.80 $\pm$ 0.18 & 1.2 $\pm$ 0.3E$-02$ & 12 & -5.50 $\pm$ 0.24 & 8.1 $\pm$ 2.9E$-03$ \\
NGC4485 & 109 & -10.80 $\pm$ 0.08 & 9.7 $\pm$ 7.2E$-02$ & 211 & -10.80 $\pm$ 0.08 & 5.1 $\pm$ 0.9E$-02$ & 102 & -7.16 $\pm$ 0.25 & 4.8 $\pm$ 0.6E$-02$ \\
NGC3738 & 40 & -9.09 $\pm$ 0.10 & 2.2 $\pm$ 0.2E$-02$ & 95 & -10.31 $\pm$ 0.25 & 1.2 $\pm$ 0.2E$-02$ & 55 & -5.52 $\pm$ 0.10 & 1.2 $\pm$ 0.2E$-02$ \\
NGC4449 & 131 & -12.63 $\pm$ 0.31 & 2.4 $\pm$ 4.4E$-01$ & 215 & -12.63 $\pm$ 0.31 & 4.1 $\pm$ 4.0E$-02$ & 84 & -5.66 $\pm$ 0.29 & 2.5 $\pm$ 0.3E$-02$ \\
NGC4656 & 68 & -13.01 $\pm$ 0.08 & 9.9 $\pm$ 1.2E$-02$ & 155 & -13.01 $\pm$ 0.08 & 5.6 $\pm$ 0.4E$-02$ & 87 & -6.76 $\pm$ 0.11 & 5.3 $\pm$ 0.4E$-02$ \\
\hline
\end{tabular}

\caption{Star Cluster properties of the LEGUS dwarf galaxies in three different age ranges. The properties listed here are: $N$ represents the number of clusters brighter than the --6~mag selection limit, \mv~represents the absolute magnitude of the brightest cluster in the $V$-band, and $CFR$ represents the cluster formation rate corrected for clusters below the completeness limit of Log(M)=3.7. The galaxies are sorted by \sfrsig~using the 10-100~Myr SFHs and the D25$\cap$HST FOV areas (see \S\ref{sec:areatest}).}
\label{tab:clustStats}

\end{table*}

\begin{figure}
  \begin{center}
  \includegraphics[scale=0.51]{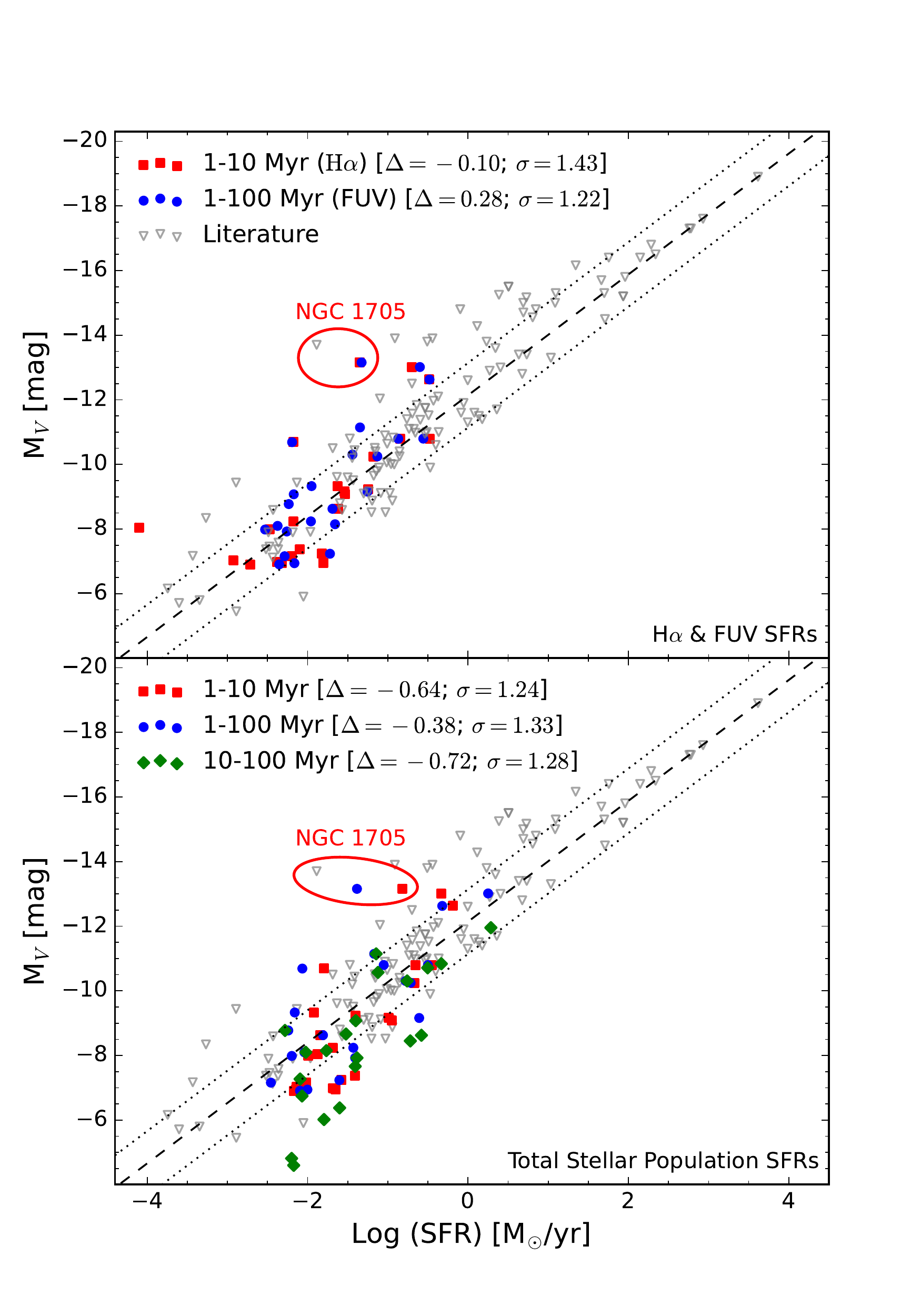}
  \caption{The brightest cluster in the $V-$band filter versus the global SFR of the LEGUS dwarf galaxies. The grey, open symbols represent data from the literature and are presented in Table~\ref{tab:lit} of Appendix~\ref{sec:append_lit}. The dashed and dotted lines in both panels represent the fit and scatter to literature data performed by \citep{weidner04}. \textit{The layout and symbols used in this figure are used in subsequent figures.} \textbf{Top:} Dust-corrected SFRs derived from \ha~and FUV fluxes: red squares and blue circles represent \ha~and GALEX FUV SFRs, respectively. The \ha~and FUV SFRs probe approximate age ranges of 1-10 and 1-100~Myr, respectively. \textbf{Bottom:} total stellar population SFRs: red squares, blue circles, and green diamonds represent age ranges of 1-10, 1-100, and 10-100~Myr, respectively.  We find that the LEGUS dwarf galaxies show good agreement with previous data and the fit to these data. The upper-left legends of both panels provide the median offset and scatter of the LEGUS dwarfs around the dashed-fit line. The red ellipses highlight the location of data for NGC~1705 which has been identified as an outlier in previous studies (see \S\ref{sec:MvSFR}).}
   \label{fig:MvSFR}
   \end{center}
\end{figure}

Figure~\ref{fig:MvSFR} shows the observed absolute $V$-band magnitude of the brightest cluster in the appropriate age range for the \ha/FUV and total stellar population SFRs. These clusters have not been corrected for internal extinction. We find good agreement in the overall normalization and scatter with previous studies which are shown as grey points. However, six of the brightest clusters in the 10-100~Myr age range fall significantly below the observed relation ($>$2~mag), while the rest show good agreement. It is not clear why the brightest clusters in these six galaxies (NGC3274, NGC4305, UGC1249, UGC685, and UGCA281) in only this age range are discrepant as the clusters do not have significant extinction values from SED fitting nor are there any anomalies or patterns in the SFHs of these galaxies.




There is also one moderate outlier brighter than the observed relationship that has also been noted in previous studies: NGC1705 \citep[e.g.,][]{whitmore00,billett02,larsen02,bastian08}. This galaxy has had a recent starburst in the last 10~Myr as shown in the SFH of \cite{Cignoni18}. Subsequently, this burst has formed a super star cluster with a spectroscopic age of $\approx$10~Myr \citep[][]{vasquez04,martins12}. This cluster is a 3.4$\sigma$ outlier in the top panel of Figure~\ref{fig:MvSFR} using the \ha~-based SFR. However, its deviation is lessened when using the 1-10~Myr total stellar population SFR (higher by an order of magnitude) in the bottom panel making it only a 2.5$\sigma$ outlier. This is in agreement with the predictions of the \cite{bastian08} model.

\subsection{Luminosity/Mass Functions} \label{sec:mfresults}
The luminosity and mass functions ($dN/dL \propto L^{\alpha}$ and $dN/dM \propto M^{\beta}$, respectively) of star clusters provide insights into cluster formation by quantifying the relative numbers of bright/massive clusters formed at a given time. Although most studies find a roughly constant power-law slope of --2 in different galaxies, there have been some findings of a trend with global star formation properties (SFR and \sfrsig) in larger samples of galaxies \citep[e.g.,][]{whitmore14a}. Here we test for trends between the LF/MF slopes of clusters against galaxy \sfrsig, and quantify the strength of possible trends via Spearman rank correlation coefficients ($\rho$). We note that the luminosity functions here are derived in units of absolute magnitudes on the Vega system as measured in the $V-$band (HST F555W or F606W) and are not converted to luminosities. Following the common practice of replacing luminosity for magnitude and fitting in logarithmic space, the exponent has been properly converted \citep[for details see,][]{Whitmore02,adamo17}.


The methodology used to construct our LFs/MFs are detailed in \cite{cook16} and C19, but we provide a brief overview here. These distributions are constructed with an equal number of clusters in each luminosity/mass bin \citep{miaz05} where the y-axis is calculated as the number of clusters per bin divided by the bin width.  We use uniform limits to fit a power law to data with luminosities brighter than the --7~mag limit for the LFs and more massive than the Log($M$/M$_{\odot})=$3.7 for the MFs. We visually verified that these limits roughly coincide with the turnovers found in the LFs/MFs of individual galaxies (see Appendix~\ref{sec:append_lfmfdndt}) and those of the composite binned clusters (see C19). 

\begin{figure}
  \begin{center}
  \includegraphics[scale=0.51]{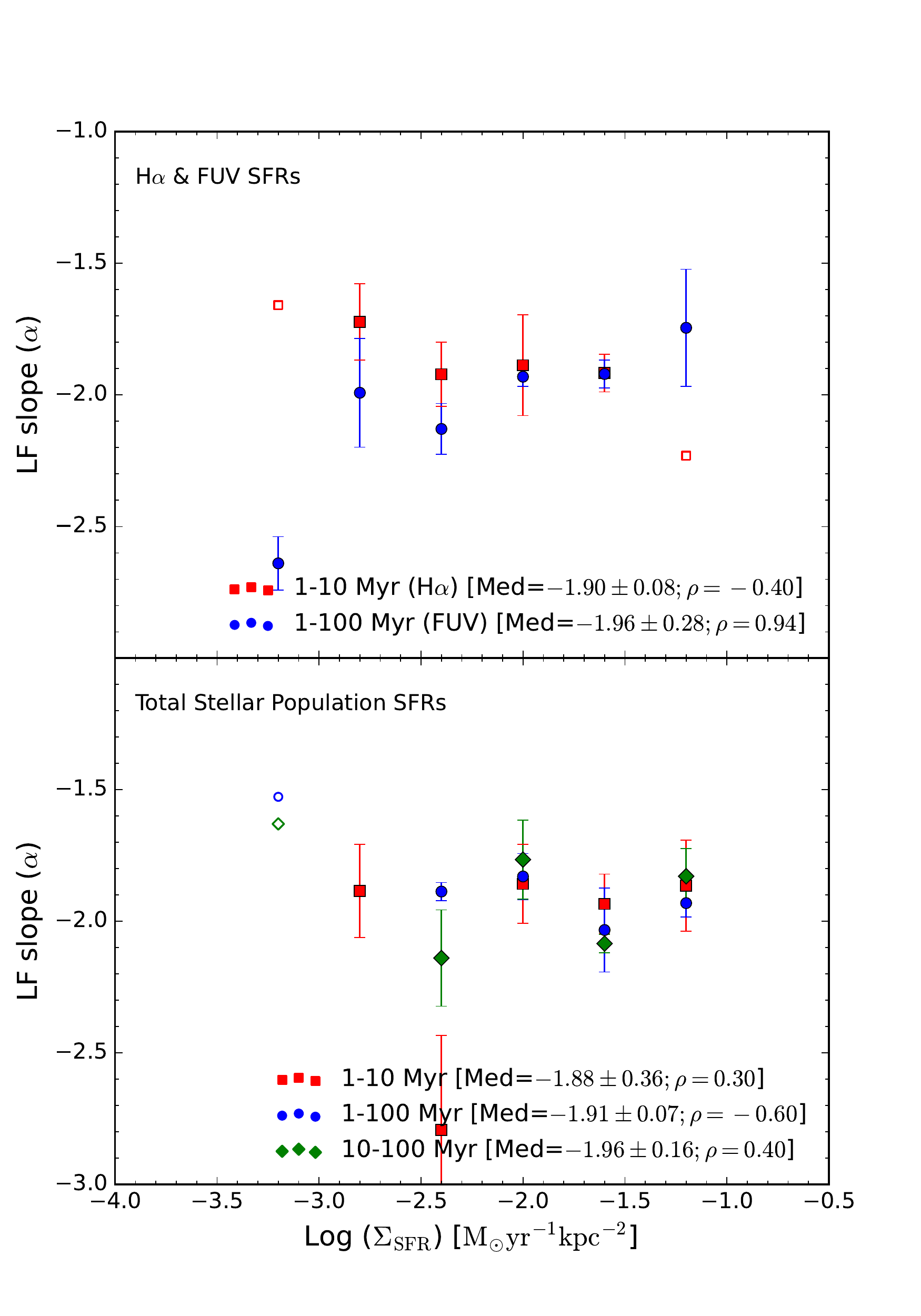}
  \caption{Luminosity function (LF) slopes binned by \sfrsig~for clusters in the LEGUS dwarf galaxies using the $V$-band filter. The panel layout and symbols are similar to those described in Figure~\ref{fig:MvSFR}. Open symbols represent bins with too few clusters \citep[N$<$30;][]{cook16} to provide a robust power-law fit and the error bars for these points have been omitted for clarity. All binned LFs were fit down to the same --7~mag limit. We find that the LFs have similar power-law slopes across \sfrsig~when using the total stellar population SFR and \ha~and FUV SFRs. The median (``Med"), standard deviation, and Spearman's rank correlation coefficient ($\rho$) for the data are presented in the legend.}
   \label{fig:LFsig}
   \end{center}
\end{figure}

\begin{figure}
  \begin{center}
  \includegraphics[scale=0.51]{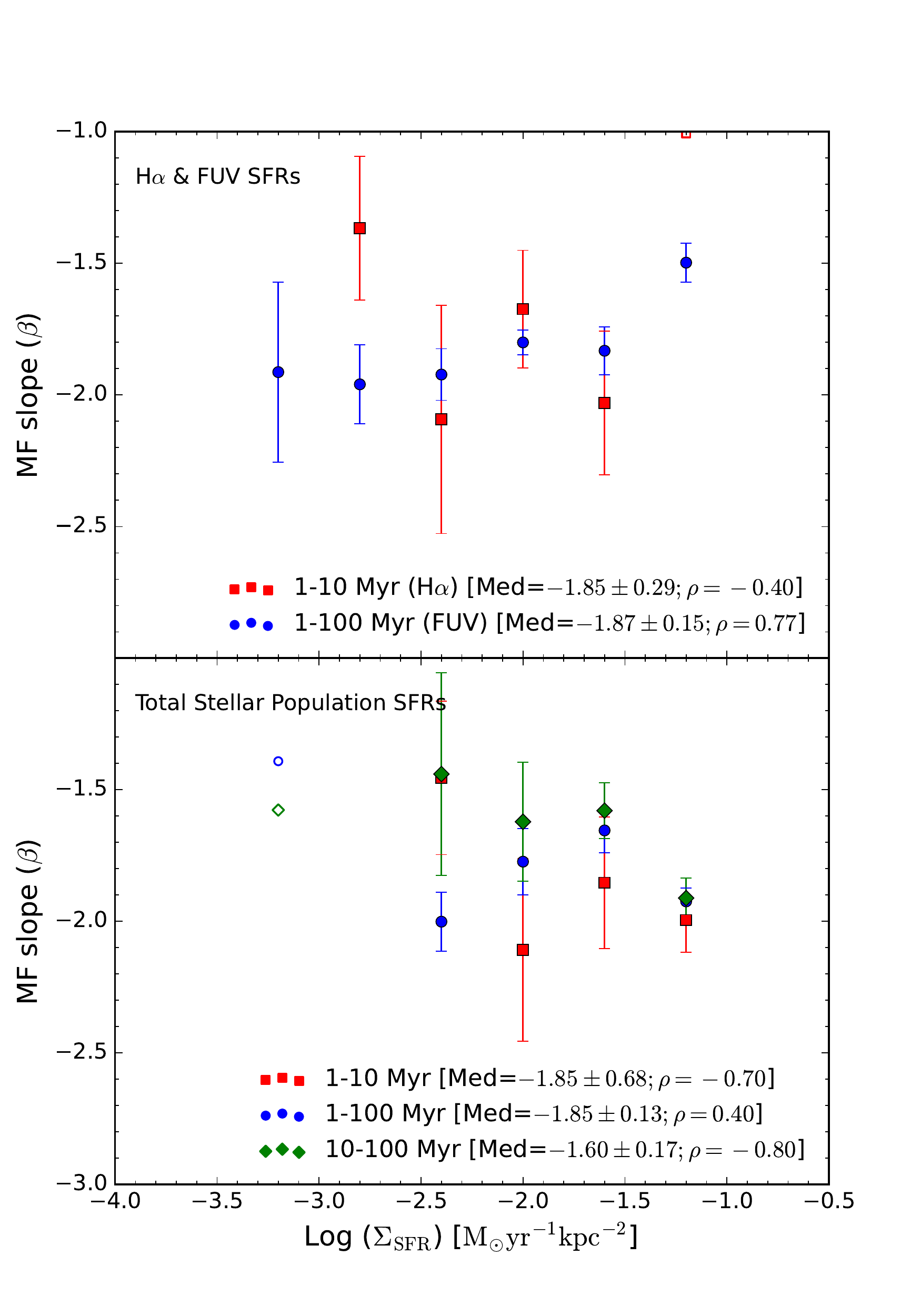}
  \caption{Mass function (MF) slopes binned by \sfrsig~for clusters in the LEGUS dwarf galaxies. The panel layout and symbols are similar to those described in Figure~\ref{fig:MvSFR}. Open symbols represent bins with too few clusters \citep[N$<$30;][]{cook16} to provide a robust power-law fit and the error bars for these points have been omitted for clarity. All binned MFs were fit down to the same Log($M$/M$_{\odot})=$3.7 limit. The MFs have similar power-law slopes with no detectable trends across \sfrsig~for both SFR methods. The median (``Med"), standard deviation, and Spearman's rank correlation coefficient ($\rho$) for the data are presented in the legend.}
   \label{fig:MFsig}
   \end{center}
\end{figure}

Figures~\ref{fig:LFsig} and \ref{fig:MFsig} show the slopes of the absolute magnitude $V-$band LFs and MFs, respectively, for cluster populations in the LEGUS dwarf galaxies versus host-galaxy \sfrsig. We note that only 5 of our galaxies have enough clusters to yield reliable power-law fits and these are shown in Appendix \ref{sec:append_lfmfdndt}. As a result, we have created composite cluster populations from different galaxies that fall into the same \sfrsig~bin to increase the cluster statistics and provide more robust LF/MF slope measurements. 

We also find that the choice of bin size by which to group different galaxies by their \sfrsig~values can affect the strength of possible correlations. An examination of the LFs/MFs and correlation coefficients for an array of bin sizes between 0.1--1.0 with steps of 0.1 reveals that bin sizes that are too small ($\la$0.2) tend to have too few galaxies ($N\approx1-2$) in each bin to provide good cluster statistics and show increased scatter in the LF/MF slopes. We also found that bin sizes that are too large ($\ga$0.6), while increasing the cluster number statistics in each bin, have the drawback of too few binned data points ($N\la$3) from which to measure a robust correlation coefficient. For the analysis of our sample, we find that bin sizes between 0.3--0.5 provide large enough bins to achieve good cluster number statistics, provide $N>3$ binned data points from which to measure robust correlations coefficient, and show largely consistent coefficients across the bin sizes. We choose the median of these bin sizes (0.4) for the remainder of the analysis when binning by global galaxy properties.

Figure~\ref{fig:LFsig} shows the binned LFs in the $V$-band (the band used to identify clusters), which do not show a significant correlation between the LF slopes and \sfrsig. Given the lack of a correlation, we report median LF slopes of $-1.90 \pm 0.08$ and $-1.96 \pm 0.28$ for the \ha\ and FUV SFRs (top panel), respectively, and a median LF slope of $-1.88 \pm 0.36$, $-1.91 \pm 0.07$, and $-1.96 \pm 0.16$ for the 1-10, 1-100, and 10-100~Myr total stellar population SFRs (bottom panel), respectively. While the 1-100~Myr cluster sample analyzed with respect to the FUV SFRs formally shows a strong correlation coefficient ($\rho\approx$0.9), this correlation relies heavily on a single data point which is made up of only 45 clusters. 

The MF slopes in Figure~\ref{fig:MFsig} also show no significant trends with \sfrsig. We report median MF slopes of $-1.85 \pm 0.29$ and $-1.87 \pm 0.15$ for the \ha\ and FUV SFRs (top panel), respectively, and a median MF slope of $-1.85 \pm 0.68$, $-1.85 \pm 0.13$, and $-1.60 \pm 0.17$ for the 1-10, 1-100, and 10-100~Myr total stellar population SFRs (bottom panel), respectively. Similar to the LFs, 1-100~Myr cluster sample using the FUV SFRs seems to show a moderate correlation coefficient ($\rho\approx$0.7), but this trend relies heavily on a single data point which is made up of only 30 clusters. 

While we have attempted to mitigate the effects of small number statistics in our sample by combining clusters across galaxies of similar \sfrsig, the cluster statistics are still too low to provide robust constraints on trends with galaxy properties as evidenced by the relatively large error bars in Figures~\ref{fig:LFsig} and \ref{fig:MFsig}. Consequently, we cannot rule out a weak trend between the luminosity/mass function slopes and galaxy star formation properties.



\subsection{Age Distributions} \label{sec:dndtresults}
The number of clusters per logarithmic age bin (i.e., the cluster age distributions; $dN/dt \propto t^{\gamma}$ is an important diagnostic for both cluster formation and dissolution. The decline of cluster numbers over time can be attributed to a combination of cluster formation and dissolution, and the slope of this distribution provides a means to quantify the effects of dissolution once the cluster formation and the presence of unbound systems have been considered. It is debated in the literature if galaxy environment plays a significant role in cluster dissolution \citep{silvavilla14,messa18b}. Here we examine if any trends exist between \sfrsig~and the cluster age distribution slopes. 

To construct these age distributions, we use similar binning procedures to those in C19 where a power-law is fit to a mass limited cluster sample with ages between 10--200~Myr. However, our fitting procedures differ in the following ways: 1) we use an expanded age range (5--500~Myr) to decrease the fitting errors and 2) we shift our age bins by half a bin so that the pile-up of clusters at Log(age/yr)$\approx$8 is not divided into two bins (see below). 

The first change is implemented based on visual inspection of the age distributions of both individual galaxies and composite clusters binned by \sfrsig. The points at 5 and 300~Myr in Figure~\ref{fig:galdndt} show a smooth continuation of the binned data in the 10--200~Myr age ranges used by C19 (for examples see Appendix~\ref{sec:append_lfmfdndt}). However, we note that there is a significant drop off in cluster numbers $\gtrsim$1 Gyr, and that these bins should not be considered in the power-law fit. The expanded age range used here requires a higher cluster mass cut so as to avoid an artificial drop in cluster numbers at ages between 200-500~Myr. Using Figure~\ref{fig:agemass} and the --7~mag completeness limit adopted in this study, we increase our mass limit cut to Log($M$/M$_{\odot})=$4 when constructing age distributions. 

The second change stems from the pile-up of cluster ages at Log(age/yr)=8.0 (see the age-mass diagram shown in Figure~\ref{fig:agemass}) which is a consequence of overlapping stellar population models near this age on a color-color diagram. Fitting cluster photometry to stellar models results in an overdensity near Log(age/yr)=8.0 and a reduction of clusters at ages just before and after this age. The complication of this overdensity arises in the placement of the age bins when fitting a power law. The bins used in C19 had endpoints on whole and half Log(age) increments (i.e., a bin size of 0.5 dex starting at Log(age/yr)=6.0) which effectively split this overdensity of clusters into two bins. By shifting the bin centers by --0.25 dex, this overdensity is now firmly in a single bin. A comparison of age distribution slopes when using shifted and non-shifted age bins shows similar overall slopes, but a decrease in the power-law fit errors by roughly 30\% when using a shifted binning procedure. In this study we utilize a shifted binning procedure to decrease the noise of age distributions and to better assess if a trend exists with \sfrsig.

\begin{figure}
  \begin{center}
  \includegraphics[scale=0.51]{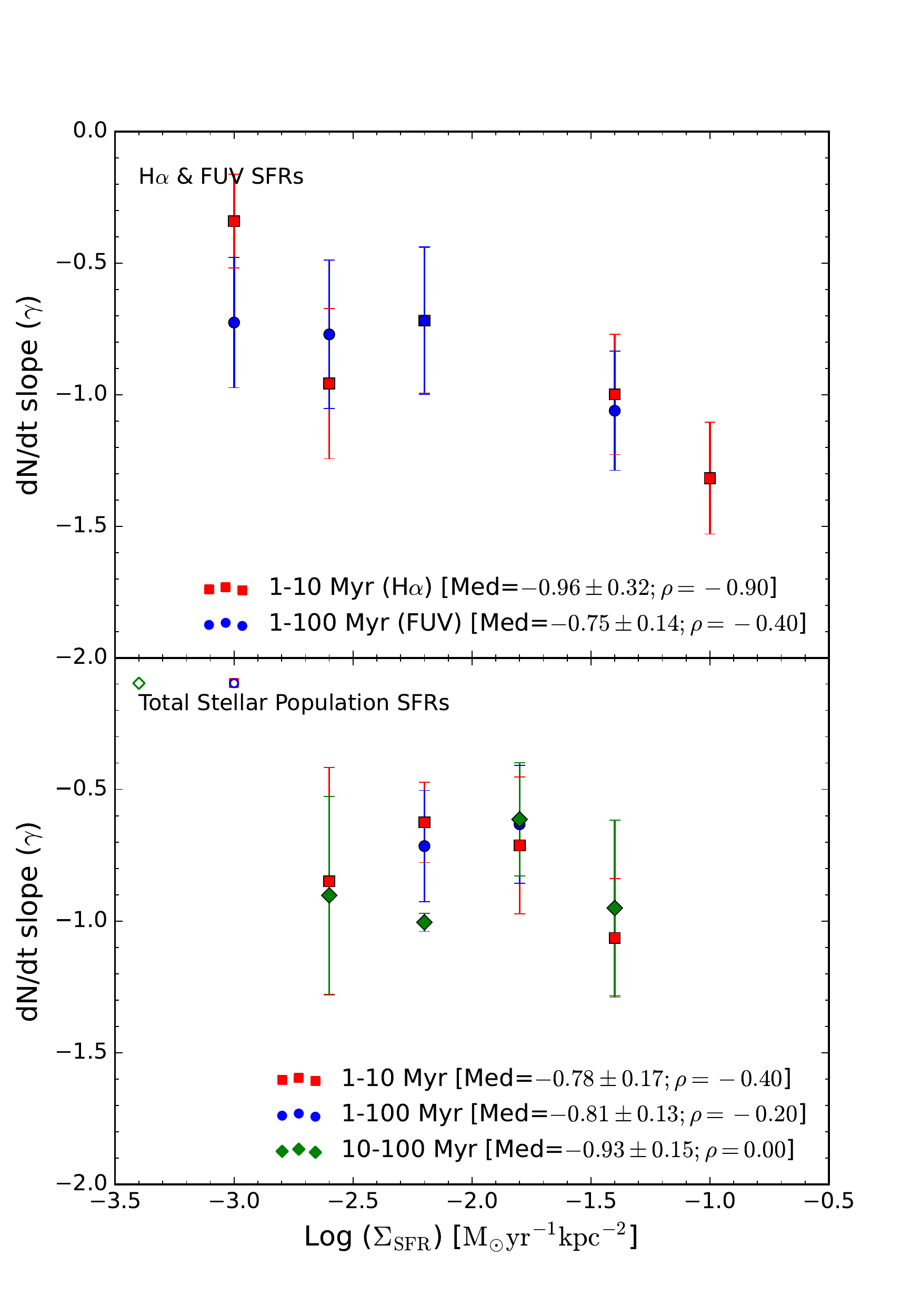}
  \caption{Age distribution slopes binned by \sfrsig~for clusters in the LEGUS dwarf galaxies. The panel layout and symbols are similar to those described in Figure~\ref{fig:MvSFR}. Open symbols represent bins with too few clusters (N$<$30) to provide a robust power-law fit and the error bars for these points have been omitted for clarity. All binned age distributions were constrained to ages between $6.7 < Log(Age/yr) < 8.7$ and cluster masses above Log($M$/M$_{\odot})=$4. The age distributions have similar power-law slopes with no detectable trends across \sfrsig~for both SFR methods. However, there is a possible steepening of the age distributions when using \ha\ SFRs, but this may not be robust as the trend relies on data points with small number statistics (i.e. lowest and highest \sfrsig data points). The median (``Med"), standard deviation, and Spearman's rank correlation coefficient ($\rho$) for the data are presented in the legend.}
   \label{fig:dndtsig}
   \end{center}
\end{figure}

Figure~\ref{fig:dndtsig} shows the age distribution slopes for cluster populations in the LEGUS dwarf and irregular sample. Similar to the LFs and MFs in the previous section, we use composite age distributions binned by \sfrsig~(bin size of 0.4) due low cluster numbers in most of the galaxy sample. We show the age distributions of the five individual galaxies with sufficient cluster numbers to compute robust power-law fits in Figure~\ref{fig:galdndt}. 

We find a weak-to-moderate trend for the \ha/FUV SFR results where a few high \sfrsig~data points tend to produce steeper age distributions while lower \sfrsig~bins show a relatively constant slope. We note here that the highest \sfrsig(\ha) data point in the top panel has only 31 clusters, thus the steep slope may be the result of random fluctuations in fitting a power law with low-number statistics. The age distributions in the bottom panel show little-to-no trend. We report median slopes of $-0.96 \pm 0.32$ and $-0.75 \pm 0.14$ for the \ha\ and FUV SFRs (top panel), respectively, and a median slope of $-0.78 \pm 0.17$, $-0.81 \pm 0.13$, and $-0.93 \pm 0.15$ for the 1-10, 1-100, and 10-100~Myr total stellar population SFRs (bottom panel), respectively. The range of median slopes measured here indicates a dissolution rate of $\approx$80\% over the past few hundred Myr, and are steeper than those typically found in nearby spiral galaxies (--0.3 to --0.2) as reported in the review of \cite{krumholz19}. These steeper slopes suggests that cluster dissolution may occur to a greater degree in dwarfs and irregular galaxies compared to spirals. 

We have also fit the age distributions over the 10--500 Myr interval, which focuses on only bound clusters. We report median slopes of $-0.86 \pm 0.23$ and $-0.84 \pm 0.28$ for the \ha\ and FUV SFRs, respectively, and a median slope of $-0.80 \pm 0.29$, $-0.77 \pm 0.24$, and $-0.88 \pm 0.13$ for the 1-10, 1-100, and 10-100~Myr total stellar population SFRs, respectively. These results are similar to those measured in the 5--500~Myr age range.


\subsection{Fraction of Stars in Clusters ($\Gamma$)} \label{sec:gamma}

The fraction of stars in gravitationally bound star clusters ($\Gamma$) is a fundamental metric of star and cluster formation. Enabled by imaging surveys of nearby galaxies with HST over the last decade, $\Gamma$ measurements across many galaxies and environments have been used to inform models of star formation \citep[e.g.,][]{kruijssen12,li18,lahen20,Dinnbier22,grudic22a}. There is currently a debate about how $\Gamma$ changes with galaxy-wide environment, where some studies (both observational and theoretical) show an increase with \sfrsig~and other studies support a constant value (see \S\ref{sec:disc} for additional discussion). The relationship between $\Gamma$ and environment will have significant impacts on our understanding of the star formation process. In this section, we present $\Gamma$ measurements for the LEGUS dwarf and irregular galaxy sample which can provide constraints at the low \sfrsig~end.

The methodology used to derive $\Gamma$ in this study is similar to those used by previous studies \citep[e.g., ][]{goddard10,adamo11c}, and the basic process is as follows: 1) we measure the raw CFR by summing the mass of clusters above our mass limit of Log($M$/M$_{\odot})=$3.7, 2) we correct for the mass of clusters below this limit by assuming a --2 power-law cluster mass function down to 100 solar masses, 3) we sum the raw and corrected cluster masses and divide by the appropriate age range (e.g., 1-10, 1-100, or 10-100~Myr). The cluster mass function correction (step 2) is computed analytically by taking the definition of the cluster mass function shown in Equation~\ref{eqn:mf} and solving for the total mass (N$\times$M) between the lower and upper mass bounds as shown in Equation~\ref{eqn:solve}:

\begin{equation}
    dN/dM = \chi \times M^{-2} \label{eqn:mf}
\end{equation}
\begin{equation}
    \begin{aligned}
    \noindent N \times M = M \times \int dN &=\chi\times \int_{M_{low}}^{M_{up}}M^{-1} dM \\
        &= \chi\times(ln[M_{up}] - ln[M_{low}]), \label{eqn:solve}
    \end{aligned}
\end{equation}

\noindent where $M_{up}$ and $M_{low}$ are the mass limit of Log($M$/M$_{\odot})=$3.7 and 2, respectively.\footnote{We note that some studies \citep[e.g.,][]{Dinnbier22} have explored a lower limit of 5M~$_{\odot}$, but there is also work that has shown that stellar groupings with masses of $\lesssim$100~M$_{\odot}$ do not survive as bound clusters \citep[e.g.][]{Grudic22b}.  Here we adopt 100~M$_{\odot}$ to facilitate comparisons with the results of most previous observational studies.}

The normalization constant $\chi$ is determined by using information from real clusters above our mass limit. Specifically, $\chi$ is determined via Equation~\ref{eqn:solve} using the total mass in clusters above the mass limit, and by setting the lower and upper limits to Log($M$/M$_{\odot})=$3.7 and twice the most massive cluster \citep{adamo15}, respectively. We note here that the multiplicative factor used to define the upper mass limit does not significantly affect the normalization nor the total correction mass in clusters below our Log($M$/M$_{\odot})=$3.7 limit. We find a difference of 0.05~dex in cluster correction mass when using factors between 1--4 times the mass of the most massive cluster.



\begin{figure}
  \begin{center}
  \includegraphics[scale=0.51]{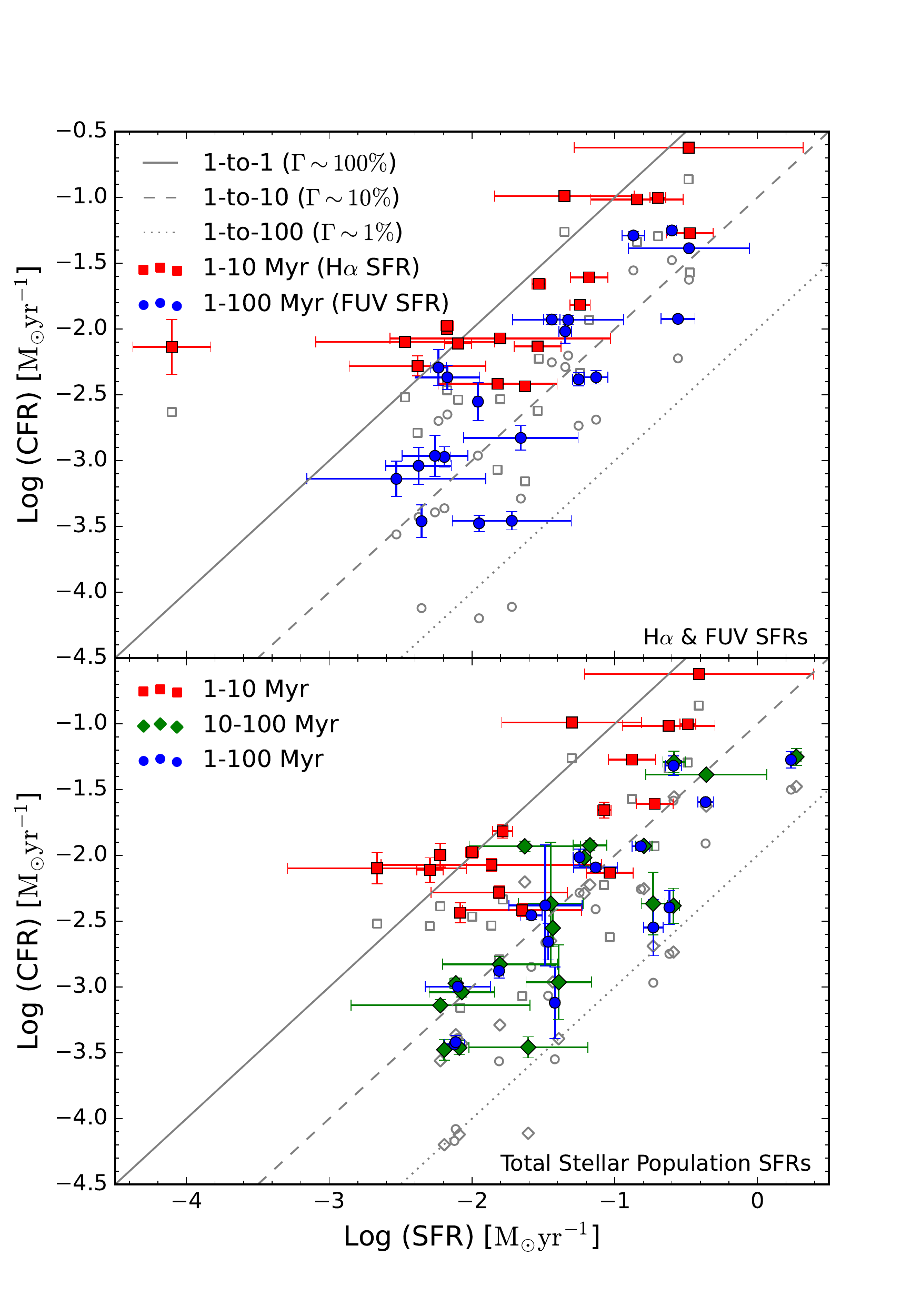}
  \caption{The cluster formation rate (CFR) vs global star formation rate (SFR). The grey, open symbols represent CFRs that have not been corrected for missing clusters with masses below the Log($M$/M$_{\odot})=$3.7 limit. The other symbol colors have the same definitions as in Figure~\ref{fig:MvSFR}. The solid, dashed, and dotted lines represent $\Gamma$ values of 100, 10, and 1\%, respectively. }
   \label{fig:SFRCFR}
   \end{center}
\end{figure}

The correction for missing clusters below our mass limit and $\Gamma$ can be visualized in a plot of the CFR as a function of the global SFR. Here we consider the CFR as the formation rate of long-lived clusters, not an initial CFR which would require corrections for cluster dissolution over time (see \S\ref{sec:GammaTime} for a discussion of the effects of cluster dissolution on our measurements). This comparison is shown in Figure~\ref{fig:SFRCFR}, where the lines indicate $\Gamma$ of 1\%, 10\% and 100\%. In this plot we show CFRs with and without corrections for clusters with masses below the detection limit of Log($M$/M$_{\odot})=$3.7 versus \ha/FUV and total stellar population SFRs in the top and bottom panels, respectively. The CFR corrections used for $\Gamma$ measurements (i.e., corrected/uncorrected CFR) is on average a factor of 2-3, where we see larger correction factors (as high as $\approx$5) for low-SFR galaxies with low numbers of clusters.

\begin{figure*}
  \begin{center}
  \includegraphics[scale=0.24]{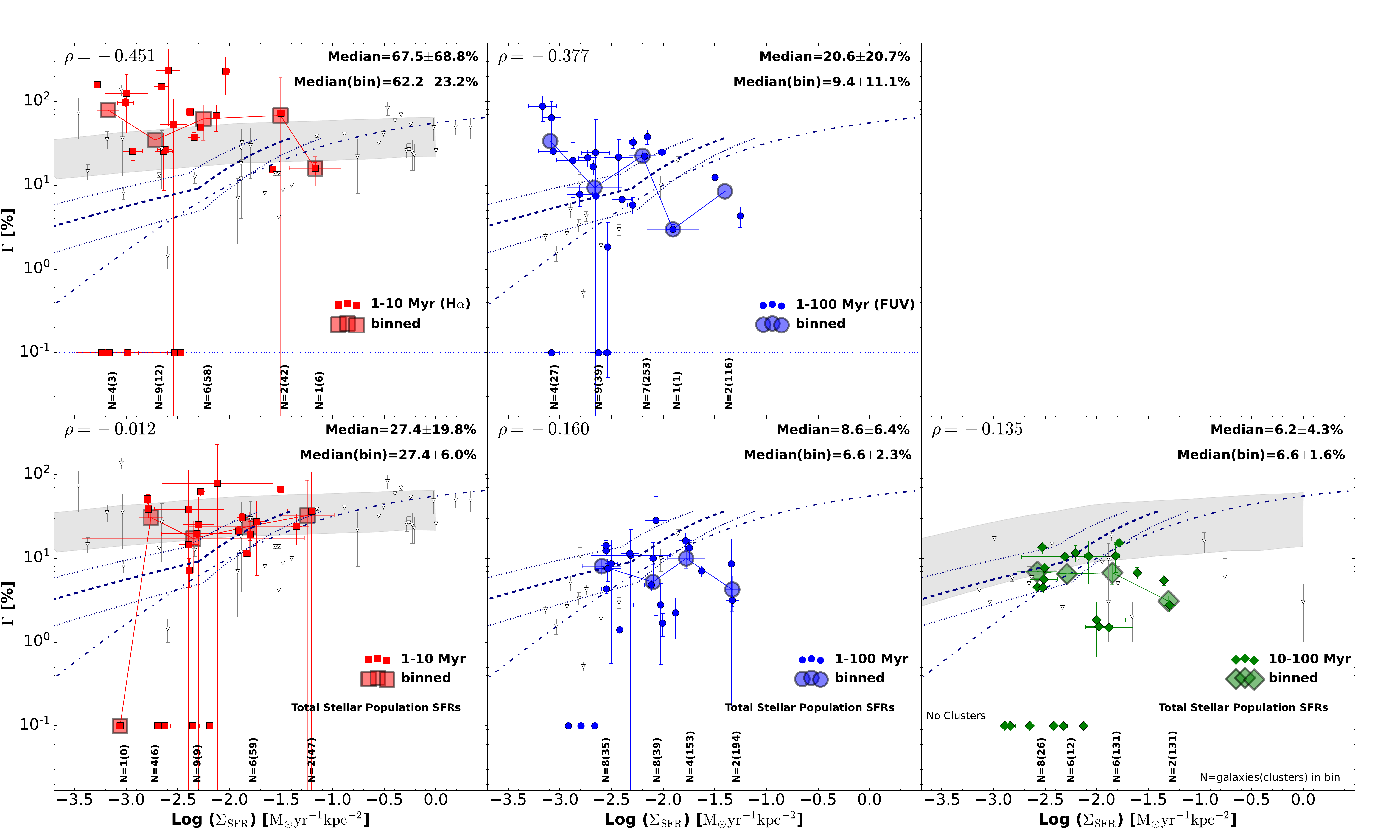}
  \caption{The fraction of stars in clusters ($\Gamma$) versus \sfrsig\ for the LEGUS dwarf galaxies. Similar to Figure~\ref{fig:MvSFR}, the top panels represent results based on integrated-light SFRs and the bottom panels represent results from SFRs based on the resolved stellar populations (stars $+$ clusters). The ages of the cluster populations used to derive $\Gamma$ have been separated for clarity. From left-to-right: red, blue, and green symbols represent results for 1-10, 1-100, and 10-100~Myr clusters, respectively. The smaller symbols represent cluster populations in individual galaxies while the larger, semi-transparent symbols represent composite clusters binned by \sfrsig. The grey points represent literature values separated into the appropriate age ranges and are tabulated in Appendix~\ref{sec:append_lit}. The median $\Gamma$ measurements for the individual galaxies (and binned by \sfrsig) studied here are presented in the upper-right corners of each panel. The horizontal dotted line represents an arbitrary $\Gamma$ value where galaxies and bins with no clusters are plotted. The dashed-dot and dashed lines are theoretical predictions derived by \citet{kruijssen12} when using the global \citep{kennicutt12} and sub-kpc \citep{bigiel08} conversion between $\Sigma_{\rm gas}$ and \sfrsig, and have been commonly used to provide a comparison to models which predict a dependence of $\Gamma$ on environment (see further discussion in Section 6.1). The shaded regions represent the age-dependent models of \citet{Dinnbier22} and have been placed in the appropriate panels. The vertical numbers at the bottom of each panel represent the number of galaxies (and clusters) in each \sfrsig\ bin. }
   \label{fig:gammasig}
   \end{center}
\end{figure*}

Figure~\ref{fig:gammasig} shows the \GS~results for individual galaxies in three age ranges commonly used in the literature: 1-10, 1-100, and 10-100~Myr. The top and bottom panels present \ha/FUV and total stellar population SFRs, respectively, that are used to calculate both $\Gamma$ and \sfrsig. Galaxies with no clusters above the completeness limit are arbitrarily shown on a horizontal line at $\Gamma$=0.1\%. The Gamma values for individual galaxies are presented in Table~\ref{tab:gammasigma} and the \sfrsig\ values can be recovered via dividing the SFRs in Tables~\ref{tab:sfh} and \ref{tab:tabsfr} by the areas listed in Table~\ref{tab:tabsfr}.

We also include theoretical comparisons for our data in Figure~\ref{fig:gammasig}. The dashed-dot and dashed lines are the predictions derived by \cite{kruijssen12} when using the global \citep{kennicutt12} and sub-kpc \citep{bigiel08} conversion between $\Sigma_{\rm gas}$ and \sfrsig, respectively. The \cite{kruijssen12} models also predict that $\Gamma$ evolves with age of the cluster population where it peaks at 3 Myr just before the onset of supernovae feedback, then slowly decreases immediately after but otherwise does not exhibits significant variation. Given the small evolution over time, we plot the peak predictions of \cite{kruijssen12} in all age panels of Figure~\ref{fig:gammasig}. We also include the models of \cite{Dinnbier22} which use N-body simulations to predict the dependence of $\Gamma$ on \sfrsig\ for different cluster population age. These models are presented as shaded regions which represent the variation of $\Gamma$ using different cluster size and mass function slope assumptions. 

We find weak correlation coefficients that range from 0.01--0.45 in absolute value for all panels of Figure~\ref{fig:gammasig}; that is, we find no significant correlation between $\Gamma$ and \sfrsig~across 2 orders of magnitude in \sfrsig~for all age ranges independent of the SFR method used. Since there is no significant trend seen in our data, we quote the median and standard deviation of $\Gamma$: 68$\pm$69\% and 21$\pm$21\% for \ha~and FUV SFRs (top panels), respectively, and 27$\pm$20\%, 9$\pm$6\%, 6$\pm$4\% for 1-10, 1-100, and 10-100~Myr (bottom panels), respectively.

We note that the literature values plotted in Figure~\ref{fig:gammasig} have been placed in the appropriate age ranges for a more consistent comparison, and have been tabulated in Appendix~\ref{sec:append_lit}. When put into appropriate age context, we find no significant correlation between $\Gamma$ and \sfrsig~in the literature with $\rho$ values that range from 0.01--0.15. We also find that the previous literature data show similar median $\Gamma$ values in our three age ranges: 27$\pm$26\%, 3$\pm$4\%, and 6$\pm$5\% for the ages 1-10, 1-100, and 10-100~Myr, respectively. The literature $\Gamma$ and \sfrsig~values are presented in Table~\ref{tab:lit} of Appendix~\ref{sec:append_lit}.

In addition to $\Gamma$~measurements of individual galaxies, we also derive composite values where clusters from multiple galaxies are binned by galaxy-wide \sfrsig~shown as larger, semi-transparent symbols in Figure~\ref{fig:gammasig}. These composite $\Gamma$ measurements are produced to mitigate low cluster numbers in many of our galaxies. We note that binned values are not taken as a simple median nor an average of individual galaxy properties in the \sfrsig~bins, but rather the binned $\Gamma$ values are computed as the ratio of the summed CFRs and the summed SFRs of the galaxies in their respective bins. Similarly, the binned \sfrsig~values are computed as the ratio of the summed SFRs and summed areas for galaxies that fall in the bin. Galaxies located in the bin with no clusters still have their SFRs and areas added to the binned $\Gamma$ and \sfrsig~values. The binned $\Gamma$ and \sfrsig\ measurements are presented in Table~\ref{tab:gammasigmabin}.

We find similar results in our binned measurements to those of individual galaxies where there is no statistically significant trend. The median $\Gamma$ values for the binned measurements are 62$\pm$23\% and 9$\pm$11\% for \ha~and FUV SFRs (top panels), respectively, and 27$\pm$6\%, 7$\pm$2\%, 7$\pm$2\% for 1-10, 1-100, and 10-100~Myr (bottom panels), respectively. The vertical text at the bottom of each panel in Figure~\ref{fig:gammasig} represent the number of galaxies (and clusters) present in the \sfrsig~bins, and can provide clues to variations seen across the \sfrsig~bins. For instance, the lowest \sfrsig~bin in the \ha-derived SFR panel is composed of 4 galaxies totaling 3 clusters above the Log($M$/M$_{\odot})=$3.7 mass limit. It is likely that several of these binned $\Gamma$ values are affected by fluctuations due to low cluster number statistics. We might also expect to see scatter for individual galaxies with low numbers of clusters.

We put our results into context of other \GS~studies that have examined cluster populations in galaxies with a range of types in \S\ref{sec:disc}, and discuss various effects that might act to increase scatter or wash out a trend between \GS.

\section{DISCUSSION} \label{sec:disc}

Two key results from the current study of star clusters in the LEGUS dwarf and irregular galaxy sample are that (1) we do not find a significant correlation in our data between the fraction of stars in a galaxy that reside in compact clusters, $\Gamma$, and the host galaxy's \sfrsig, and (2) $\Gamma$ decreases with the age of the population.  Our work provides the largest sample of homogeneously derived $\Gamma$ and \sfrsig~values for dwarf and irregular galaxies available to date. Our study also has the advantage of having SFRs derived from counts of young stars and stellar clusters (``total stellar population SFRs''), a method which complements more commonly used SFR indicators such as the integrated \ha~or FUV flux measurements, which can be problematic for galaxies with low rates of star formation.  Throughout this work, we have been careful to plot, compare, and quantify results for $\Gamma$ between galaxies measured over the same interval of age, whether the results were determined here or published in the literature. Another novel aspect of this work is that we combine the clusters across different galaxies with similar \sfrsig~properties to increase number statistics for computing $\Gamma$.

\subsection{The Relationship Between $\Gamma$ and \sfrsig}

We start with a summary of previous results.  \citet{larsenrichtler00} were the first to (indirectly) measure a positive correlation in the \GS~relation, finding that higher \sfrsig~environments tend to have a higher fraction of $U$-band luminosity in star clusters than found in lower \sfrsig~environments.  Nearly a decade later, \citet{goddard10} determined $\Gamma$ for clusters in NGC3256 and added measurements from the literature for other galaxies (Antennae, LMC, NGC 1569, NGC 6946, MW, M83, SMC) to find a positive correlation between $\Gamma$ and \sfrsig.  Subsequent observational studies supported this picture \citep[e.g.,][and references therein]{adamo11c,cook12,adamo15, Matthews18,krumholz19}, suggesting that higher \sfrsig\ (and $\Sigma_{gas}$) environments can reach higher star formation efficiencies (SFEs) to produce an increased fraction of bound star clusters.

Theoretical models provided predictions generally consistent with these observations. \cite{kruijssen12} developed a theoretical framework for $\Gamma$ based on the density distribution of the ISM, which showed that an increased fraction of bound clusters is expected to be produced at higher SFEs in higher density ISM conditions due to shorter free-fall times. Subsequently both semi-analytic \citep{li18} and cosmological galaxy simulations \citep{pfeffer19,lahen20} found a similar increasing \GS~relationship, driven by the underlying gas pressure and SFE of the ISM. 

Other observational work has indicated that the observed \GS~relation may be due to a subtle bias arising from $\Gamma$ measurements determined from clusters with different ages in different galaxies.  \cite{chandar17} separated the compiled $\Gamma$ values from \cite{adamo15} for 22 galaxies by the cluster age interval to make this comparison.  They found that $\Gamma$ values determined for galaxies with higher \sfrsig~(Log(\sfrsig) $>-2$) were based on 1-10~Myr cluster populations, while nearly all of those with lower \sfrsig~values were based on older (10-100 or 1-100~Myr) clusters. When they compared $\Gamma$ values determined from similar age cluster populations in 8 galaxies (including irregulars, dwarf starbursts, spirals, and mergers) that span 4 orders of magnitude in \sfrsig, they found no trend of $\Gamma$ with \sfrsig.  However, they found that $\Gamma$ decreased with age, by comparing results in the 1-10, 10-100, and 100-400~Myr intervals, and attributed this result to early cluster dissolution.

The more recent models of \cite{Dinnbier22} are consistent with the observational results of \cite{chandar17} for young clusters (i.e., \Glow$\sim 30\%$ with no dependence on \sfrsig), and also show that $\Gamma$ decreases with age.  However, their values of $\Gamma$ for older clusters are a factor of 10 higher when compared with \cite{chandar17}.  
While the \cite{Dinnbier22} models at older ages do predict a dependence of $\Gamma$ on \sfrsig, this dependence is much weaker than the models of \cite{kruijssen12}. The \cite{Dinnbier22} simulations use the NBODY6 code \citep{Aarseth03} and assumes a $\Gamma$ of 100\% at birth and allows the clusters to evolve over time given mass loss from gas expulsion and tidal effects from the host galaxy. However, \cite{Dinnbier22} suggest that additional sources of cluster disruption (i.e., interactions with giant molecular clouds) may be needed in their models to more accurately reproduce observational data.

In this work, we further explore $\Gamma$ in the low \sfrsig~regime. Our star cluster sample aggregated from 23 dwarf and irregular galaxies is the largest to date, and covers the range of Log(\sfrsig) values from -3 to -1.3, with a median of -2.3.  We do not find any statistically significant trends between $\Gamma$ and \sfrsig\ when we compare results within matched age intervals. This is true for both the analysis based on $\Gamma$ measurements for individual galaxies, as well as when $\Gamma$ is computed for the combined populations of clusters in bins of \sfrsig. Our homogeneously determined $\Gamma$ measurements and fairly small range in distance ($\approx3-13$ Mpc), should result in lower systematic uncertainties compared with other previous studies \citep[which included clusters out to $\approx$80 Mpc][]{adamo11c,chandar17}.

While our data do not show a trend over the nearly factor of 100 range in \sfrsig\ covered in this work, we cannot rule out a weak trend in \GS~or an increase in $\Gamma$ at higher values of \sfrsig. In particular, a trend of increasing $\Gamma$ may exist if these data are combined with more extreme \sfrsig~environments such as those observed in BCDs and merger systems \citep{adamo11c}. On-going and future studies which explore these extremes, for example from the HIPEEC \citep{adamo20} and CCDG (HST GO-15649; PI: Chandar) projects, combined with the larger sample of spiral galaxies in the LEGUS \citep[][Adamo et al. in prep]{calzetti15} and PHANGS-HST \citep{Lee21} surveys, are needed to cover the full range of \sfrsig~found in the nearby Universe. When comparing our data to the predictions of theoretical models, we find that our \Glow\ values are in better agreement with those from \cite{Dinnbier22}, but that \Gmid\ values are in nominal agreement with both \cite{kruijssen12} and \cite{Dinnbier22}.  To infer conclusions about the nature of cluster formation from such comparisons, the underlying assumptions in calculation of $\Gamma$ and star formation rates in both the observational data and theoretical predictions must be carefully examined for consistency.

\subsection{The Evolution of $\Gamma$ Over Time}\label{sec:GammaTime}

Our study of dwarf and irregular galaxies provides new constraints on the evolution of $\Gamma$ over the first 100~Myr at low \sfrsig. In Section~5.4, we found that $\Gamma$ decreases over time starting at $27\pm6$\% for the binned results ($27\pm20$\% for individual galaxies) in the first 10~Myr, then drops to $7\pm2$\% ($6\pm4$\% for individual galaxies) over the 10-100~Myr interval. These values for low \sfrsig\ galaxies are similar to the results found by \cite{chandar17} for a sample of 8 galaxies where they report median $\Gamma$ values of $27\pm$9\% and $5\pm2.5$\% for 1-10 and 10-100~Myr, respectively. The consistency is notable because the \citet{chandar17} sample covers higher values of and a larger range of \sfrsig~and the galaxies are at very different distances.

The drop in the fraction of stars found in clusters over time is likely due to the dissolution of the clusters rather than to changes in their formation rate, since we have independently measured the star formation rate in each galaxy matched to the age interval from resolved stellar population work \citep{cignoni19}. Another indication that clusters dissolve over time comes from the age distribution of the clusters which are presented in \S\ref{sec:dndtresults}. The fitted negative power-law indices for our ensemble cluster population (Figure~\ref{fig:dndtsig}) show no obvious trend with \sfrsig\ and indicate an upper limit of $\approx$80\% dissolution over the past 500~Myr \citep[see also,][]{cook19a,krumholz19,whitmore20}.

However, in order to quantify the degree of dissolution, the fraction of unbound systems in cluster samples must also be constrained and this has been a major challenge for cluster evolution studies. A common approach to identifying bound clusters is by morphological selection; i.e. identifying objects which are compact, single peaked, and centrally concentrated (class 1 and 2 in our study).  While such morphological selections increase the probability that a given cluster is bound, the classification by itself is an insufficient condition for establishing boundedness, particularly at the youngest ages \citep{krumholz19}.

The recent study of \cite{Brown21} has attempted to constrain the fraction of unbound star clusters in 31 LEGUS galaxies (including 18 galaxies in this analysis).  \cite{Brown21} compared the instantaneous crossing times (based on sizes derived from profile fitting) and cluster ages. They find that most clusters with ages $\gtrsim$10~Myr are older than their crossing times. In addition, if we adopt the values in Figure 15 from \cite{Brown21} (see below for why this may be oversimplified due to the presence of other dissolution mechanisms), clusters with ages $\lesssim$10~Myr contain roughly a 50\% mix of those which are both older and younger than their crossing times. Taking this contamination into account would reduce our dissolution factor for Gamma by an equivalent factor of roughly 2. Hence, when including uncertainties, it becomes difficult to unequivocally state that dissolution is taking place over the first 10 Myr and that the change in Gamma from 1-10 Myr and 10-100 Myr is due to dissolution in the dwarfs analyzed in this work. In future work, it would be interesting to attempt to account for unbound clusters in studies of cluster dissolution as most clusters continuously lose mass to some degree over their lifetime. Consequently, it would be important to account for older class 1+2 clusters which are becoming unbound due to various other processes \citep[e.g., GMC interactions, tidal shear, 2-body relaxation;][]{spitzer58,bastian12a,fall12} for a robust measurement of the dissolution rate. It is this degree of "contamination" from unbound clusters in different age ranges that makes such studies a true challenge, and limits most measurements to be upper limits.

We end this section by noting a key open issue: the value of $\Gamma$ at birth. That is, are all stars born in clusters?  Our focus in this paper is on providing measurements of $\Gamma$ at the times after birth between 1 and 100 Myr for which our UV-optical HST data can provide constraints.  While such measurements can be compared to models to infer Gamma at birth \citep[e.g.][]{Dinnbier22}, a new generation of observational studies are imminent with JWST, which provides the resolution and sensitivity in the infrared to enable a census of embedded star clusters in galaxies beyond the Magellanic Clouds and measurement of $\Gamma$ at birth throughout a full range of galactic environments \citep{Lee21}.  There is much to look forward to in this field.

\subsection{Sources of Uncertainty and Physical Scales} \label{sec:discunc}

Sources of uncertainty that could affect our estimates of $\Gamma$ and \sfrsig, involve low number statistics in our star cluster samples, and errors in our age (and hence mass) estimates. In addition, the choice of the physical scales over which $\Gamma$ and \sfrsig~are measured could affect the nature of the observed relationship.



A major source of uncertainty arises from the low number of clusters in individual dwarf and low-mass, irregular galaxies, a challenge inherent to these kinds of studies. Small number statistics leads to increased scatter in $\Gamma$ measurements due to sampling of the cluster mass function \citep[e.g.,][]{cook12}. This large scatter can be seen in our sample where the standard deviation of $\Gamma$ measurements for individual galaxies is typically $\approx$70\% of the median value. To address this issue, we binned the cluster populations of our galaxies by \sfrsig.  This strategy reduced the scatter around the median $\Gamma$ values down to $\approx$25\%, and still did not reveal a correlation between \GS. Future work to further increase the total number of clusters which contribute to each \sfrsig\ bin plotted in Figure~12 are needed to improve our estimates of $\Gamma$.

Another potential source of uncertainty in our results comes from inaccurate estimates of cluster ages, and hence masses, of which there can be two causes: 1) low-mass clusters for which the stellar IMF is not fully sampled, and are not well modeled by standard population synthesis simple stellar (single-aged; mono-metallicity) populations which are based on $\gtrsim10^{5}~M_{\odot}$ ``units" and 2) age-extinction degeneracy. Age estimates based on deterministic models as used here can have a significant impact on individual clusters with masses below $\approx10^{3.5}~M_{\odot}$ \citep[][]{fouesneau12,krumholz15} since the upper portion of the stellar IMF is not fully populated. We mitigate this issue by using a cluster mass limit of $M>10^{3.7}$ for our $\Gamma$ measurements. Therefore, while uncertainties in age dating due to IMF sampling effects will affect age and mass estimates for individual clusters, they are unlikely to significantly impact the overall $\Gamma$ results.


Uncertainties in the age estimates arising from the age-extinction degeneracy must also be considered, because they can potentially shift clusters between the age bins used in our analysis. For example, clusters with ages between 10 and 100 Myr with little reddening have optical broad-band properties similar to those with ages between 1 and 10 Myr and moderate reddening.  If this type of incorrect age-dating is widespread, it could artificially increase $\Gamma$ within the $1-10$~Myr age range and decrease $\Gamma$ measured for the $10-100$~Myr age range.  One way these issues can be prevented is by breaking the degeneracy with additional data, for example including narrow-band photometry for a hydrogen line like H$\alpha$ directly in the spectral energy distribution fitting \citep[e.g.,][]{fall05,chandar10a, whitmore20}.

We can assess the potential impact by examining issues in NGC~4449, for which \citet{whitmore20} has derived improved cluster ages based on the addition of HST H$\alpha$ narrowband photometry.  Moreover, NGC~4449 provides the largest number of clusters of any individual galaxy in the current analysis. Based on a comparison between the improved ages derived by \citet{whitmore20} and those used here, we find that ages in the 10-100~Myr interval are relatively robust to possible effects from the age-extinction degeneracy.  In fact, the NGC~4449 clusters that have estimated ages younger than 10~Myr and E(B-V)$>0.6$~mag in this work, {\em nearly all have revised ages older than 100~Myr} in \cite{whitmore20}; see their study for a more detailed discussion on the effects of age-extinction degeneracy. Consequently, if the results of \citet{whitmore20} can be generalized to other galaxies in our sample, then the the youngest point (\Glow) in the $\Gamma$ versus age analyses will be artificially elevated. To quantify the overestimate in \Glow\ would require a systematic reanalysis of the ages for the clusters in our LEGUS dwarf and irregular sample, and should be investigated in future work.  Nevertheless it should be noted that the age distributions from \citet{whitmore20} yield slopes consistent with those reported here within the uncertainties. 


Assuming that the uncertainties discussed above should affect all galaxies similarly to NGC~4449, there should be no change in our conclusion that there is a lack of correlation between $\Gamma$ and \sfrsig\ on galaxy scales in our data, although it remains possible that a weak trend exists. Again, better statistics from including more clusters and increasing the range of \sfrsig\ are needed before weak trends can be ruled out.

Finally, it is also possible a \GS\ correlation exists on smaller (sub-kpc) physical scales \citep[e.g.,][]{adamo15,johnson16,whitmore20} that better match the local environments and GMCs out of which bound clusters form \citep[e.g., high gas pressures;][]{Hunter18}. The theoretical and simulation work discussed earlier showed that the local variation in gas pressure and SFE has a significant effect on $\Gamma$, where a scatter of $\approx$0.25~dex is expected \citep{kruijssen12,li18} and may be higher in low \sfrsig~environments \citep{pfeffer19}. In particular, the high-resolution (sub-parsec) simulations of dwarf galaxies performed by \cite{Hislop22} indicate that $\Gamma$ can vary by an order of magnitude when changing the SFE. We will investigate $\Gamma$ on sub-kpc scales in our dwarf and irregular galaxies by binning by local \sfrsig~environment in an upcoming work.

\begin{table*}

{\textbf{Fraction of Stars in Clusters for Individual Galaxies}}\\
\begin{tabular}{lccccc}

\hline
\hline
Galaxy         & $\Gamma$ &  $\Gamma$  &  $\Gamma$   & $\Gamma$    & $\Gamma$      \\ 
Name           & (Ha)     &  (FUV)     &  (1-10 Myr) & (1-100 Myr) & (10-100 Myr)  \\ 
               & (\%)     &  (\%)      &  (\%)       & (\%)        & (\%)          \\ 
\hline

UGC7408 & \ldots & 87.5 $\pm$ 29.4 & \ldots & \ldots & \ldots \\
UGC5139 & \ldots & \ldots & \ldots & \ldots & \ldots \\
UGC4459 & \ldots & \ldots & \ldots & \ldots & \ldots \\
UGC4305 & \ldots & \ldots & \ldots & \ldots & \ldots \\
NGC5238 & 150.4 $\pm$ 11.5 & 16.6 $\pm$ 3.4 & 62.7 $\pm$ 7.0 & 12.4 $\pm$ 1.5 & 4.5 $\pm$ 0.9 \\
UGC5340 & 26.7 $\pm$ 4.4 & 21.3 $\pm$ 5.0 & 38.7 $\pm$ 8.4 & 14.2 $\pm$ 2.2 & 13.6 $\pm$ 2.2 \\
UGC7242 & \ldots & 7.8 $\pm$ 2.2 & \ldots & 4.3 $\pm$ 0.5 & 4.5 $\pm$ 0.5 \\
IC559 & 158.1 $\pm$ 19.7 & 25.5 $\pm$ 8.6 & 51.3 $\pm$ 6.5 & 7.6 $\pm$ 2.2 & 5.6 $\pm$ 1.7 \\
ESO486G021 & 53.7 $\pm$ 95.5 & 6.8 $\pm$ 6.4 & 38.3 $\pm$ 74.3 & 8.6 $\pm$ 8.1 & 7.7 $\pm$ 1.1 \\
NGC5477 & 25.4 $\pm$ 24.4 & 1.8 $\pm$ 1.8 & 14.6 $\pm$ 14.3 & 1.4 $\pm$ 1.4 & \ldots \\
UGC685 & 235.2 $\pm$ 339.6 & 24.6 $\pm$ 36.3 & 78.7 $\pm$ 149.5 & 11.4 $\pm$ 16.7 & \ldots \\
NGC4248 & 97.0 $\pm$ 21.3 & 63.9 $\pm$ 36.0 & 19.8 $\pm$ 16.1 & 10.9 $\pm$ 11.2 & 10.5 $\pm$ 11.8 \\
NGC1705 & 230.7 $\pm$ 260.0 & 24.9 $\pm$ 22.4 & 67.2 $\pm$ 87.3 & 28.4 $\pm$ 26.3 & 11.6 $\pm$ 2.6 \\
UGCA281 & 15.6 $\pm$ 0.3 & 3.0 $\pm$ 0.4 & 30.7 $\pm$ 3.7 & 4.8 $\pm$ 0.7 & \ldots \\
IC4247 & \ldots & 21.6 $\pm$ 13.4 & \ldots & 10.0 $\pm$ 5.4 & 10.6 $\pm$ 5.7 \\
UGC695 & 125.9 $\pm$ 140.5 & 19.8 $\pm$ 12.7 & 25.2 $\pm$ 29.1 & 2.8 $\pm$ 2.2 & 1.8 $\pm$ 1.2 \\
UGC1249 & 25.5 $\pm$ 9.7 & 7.4 $\pm$ 1.1 & 7.2 $\pm$ 2.8 & 1.7 $\pm$ 0.5 & 1.5 $\pm$ 0.5 \\
NGC3274 & 37.3 $\pm$ 11.2 & 5.8 $\pm$ 1.3 & 11.5 $\pm$ 3.5 & 2.2 $\pm$ 1.1 & 1.5 $\pm$ 0.8 \\
NGC5253 & 16.0 $\pm$ 6.1 & 4.3 $\pm$ 1.2 & 24.2 $\pm$ 9.7 & 13.4 $\pm$ 3.8 & 10.7 $\pm$ 3.9 \\
NGC4485 & 67.5 $\pm$ 50.4 & 38.0 $\pm$ 7.4 & 27.4 $\pm$ 21.1 & 16.2 $\pm$ 3.7 & 15.3 $\pm$ 3.1 \\
NGC3738 & 75.0 $\pm$ 9.0 & 32.6 $\pm$ 5.1 & 19.5 $\pm$ 3.2 & 7.1 $\pm$ 1.0 & 6.7 $\pm$ 1.1 \\
NGC4449 & 72.5 $\pm$ 134.0 & 12.4 $\pm$ 12.1 & 36.6 $\pm$ 70.3 & 8.6 $\pm$ 8.4 & 5.5 $\pm$ 0.7 \\
NGC4656 & 49.5 $\pm$ 6.2 & 22.3 $\pm$ 2.0 & 21.2 $\pm$ 2.8 & 3.1 $\pm$ 0.5 & 2.8 $\pm$ 0.5 \\

\hline
\hline
Median & 67.5 $\pm$ 68.8   & 20.6 $\pm$ 20.7   & 27.4 $\pm$ 19.8   & 8.6 $\pm$ 6.4   & 6.2 $\pm$ 4.3 \\ 
\hline
\end{tabular}
\caption{The cluster formation efficiency ($\Gamma$) measurements for individual galaxies in our dwarf and irregular galaxy sample. The columns present these data for the different SFR indicators and age ranges. The corresponding \sfrsig~values can be computed by dividing the SFRs from Tables~\ref{tab:sfh} and \ref{tab:tabsfr} by the areas from Table~\ref{tab:tabsfr}. The bottom line presents the median and standard deviation of $\Gamma$ values for the different SFR indicators and age ranges. }

\label{tab:gammasigma}

\end{table*}

\begin{table*}
{\textbf{Fraction of Stars in Clusters Binned by \sfrsig}}\\ 
\begin{tabular}{cccccc}

\hline
\hline
SFR Method        & \sfrsig   & \sfrsig                          & $\Gamma$   & N galaxies  & N clusters  \\ 
(Tracer/Age range)& bin range & in bin                           & in bin     & in bin      & in bin      \\ 
                  &           & ($M_{\star} yr^{-1} kpc^{-2}$)   & (\%)       & (N)         & (N)         \\ 
\hline
\ha        & -3.5 -- -3.0  & -3.17 $\pm$ 0.11  & 79.0 $\pm$ 11.3  &  4  &  3  \\ 
           & -3.0 -- -2.5  & -2.72 $\pm$ 0.08  & 34.6 $\pm$ 16.3  &  9  &  12  \\ 
           & -2.5 -- -2.0  & -2.25 $\pm$ 0.04  & 62.2 $\pm$ 27.8  &  6  &  58  \\ 
           & -2.0 -- -1.5  & -1.51 $\pm$ 0.01  & 68.0 $\pm$ 125.0  &  2  &  42  \\ 
           & -1.5 -- -1.0  & -1.17 $\pm$ 0.25  & 16.0 $\pm$ 6.1  &  1  &  6  \\ 
FUV        & -3.5 -- -3.0  & -3.09 $\pm$ 0.23  & 33.6 $\pm$ 11.9  &  4  &  27  \\ 
           & -3.0 -- -2.5  & -2.67 $\pm$ 0.26  & 9.4 $\pm$ 1.9  &  9  &  39  \\ 
           & -2.5 -- -2.0  & -2.20 $\pm$ 0.13  & 22.4 $\pm$ 3.3  &  7  &  253  \\ 
           & -2.0 -- -1.5  & -1.90 $\pm$ 0.25  & 3.0 $\pm$ 0.4  &  1  &  1  \\ 
           & -1.5 -- -1.0  & -1.40 $\pm$ 0.01  & 8.5 $\pm$ 6.6  &  2  &  116  \\ 
1-10 Myr   & -3.5 -- -3.0  & -3.06 $\pm$ 0.25  & \ldots  &  1  &  0  \\ 
           & -3.0 -- -2.5  & -2.76 $\pm$ 0.12  & 30.7 $\pm$ 5.0  &  4  &  6  \\ 
           & -2.5 -- -2.0  & -2.35 $\pm$ 1.08  & 17.3 $\pm$ 10.5  &  9  &  9  \\ 
           & -2.0 -- -1.5  & -1.81 $\pm$ 0.58  & 24.1 $\pm$ 11.4  &  6  &  59  \\ 
           & -1.5 -- -1.0  & -1.24 $\pm$ 0.26  & 32.6 $\pm$ 52.4  &  2  &  47  \\ 
1-100 Myr  & -3.0 -- -2.5  & -2.59 $\pm$ 0.19  & 8.1 $\pm$ 1.7  &  8  &  35  \\ 
           & -2.5 -- -2.0  & -2.10 $\pm$ 0.45  & 5.2 $\pm$ 3.2  &  8  &  39  \\ 
           & -2.0 -- -1.5  & -1.78 $\pm$ 0.18  & 10.0 $\pm$ 2.4  &  4  &  153  \\ 
           & -1.5 -- -1.0  & -1.33 $\pm$ 0.07  & 4.2 $\pm$ 1.9  &  2  &  194  \\ 
10-100 Myr & -3.0 -- -2.5  & -2.58 $\pm$ 0.19  & 6.9 $\pm$ 1.2  &  8  &  26  \\ 
           & -2.5 -- -2.0  & -2.29 $\pm$ 0.43  & 6.5 $\pm$ 3.6  &  6  &  12  \\ 
           & -2.0 -- -1.5  & -1.85 $\pm$ 0.32  & 6.7 $\pm$ 1.9  &  6  &  131  \\ 
           & -1.5 -- -1.0  & -1.31 $\pm$ 0.07  & 3.1 $\pm$ 0.4  &  2  &  131  \\ 

\hline
\end{tabular}

\caption{The cluster formation efficiency ($\Gamma$) measurements for cluster combined across galaxies whose \sfrsig~values fall in the same range. The $\Gamma$ and \sfrsig~measurements in the bins were computed as if the clusters came from a single galaxy. The first column indicates what SFR indicator and age range was used to compute the binned measurements.}
\label{tab:gammasigmabin}

\end{table*}





\section{SUMMARY}

In this study we examine the following relationships between the ensemble properties of star clusters and global star formation properties of their host galaxies:

\begin{itemize}
    \item $M_{V}^{brightest}-$SFR: the absolute magnitude of the brightest cluster as a function of the integrated galaxy SFR.
    \item The power-law index of the star cluster luminosity function (dN/dL $\propto L^{\alpha}$), mass function (dN/dM $\propto L^{\beta}$), and age distribution (dN/dt $\propto L^{\gamma}$) versus \sfrsig\ (SFR/Area).
    \item \GS: the fraction of stars in clusters ($\Gamma$) as a function \sfrsig.
\end{itemize}

This analysis is performed on a sample of 23 dwarf and irregular galaxies in the LEGUS survey that span a Log(\sfrsig) from --3 to --1.3 $\sigunit$, i.e., at the low end of star formation densities in the nearby universe. Our sample contains a total of 1371 clusters (class 1 and 2), where 436 are younger than 100~Myr and above our mass limit cut (Log($M$/M$_{\odot})>$3.7). However, the majority of our galaxies contain very few clusters with a median 21 clusters per galaxy at all ages; 70\% of the clusters are from 4 galaxies: NGC~4449, NGC~4485, NGC~4656, and NGC~3738.


A key strength of this analysis is the availability of global SFRs measured from temporally resolved star formation histories (SFHs) which provide the means to match both cluster and host-galaxy properties in distinct intervals (1-10, 1-100, and 10-100~Myr). We also compute SFRs from integrated \ha~and UV luminosities that roughly probe 1-10 and 1-100~Myr old stellar populations, respectively. A comparison of SFRs on similar timescales shows good agreement with moderate scatter, but the \ha~SFRs are offset by a factor of 2 at lower SFRs as discussed extensively in previous work \citep[e.g.,][]{lee09b,meurer09,pflam09,weisz12}.




When examining the relationships between star clusters and host-galaxy properties, we find a trend between the magnitude of the brightest cluster and the host galaxy SFR consistent with previous work \citep{whitmore00,larsen02,weidner04,bastian08,adamo11c,cook12,Randriamanakoto13,whitmore14a}.  However, we find no significant trends between the power-law indices of the luminosity and mass functions versus \sfrsig, which suggests that galaxy environment may not play a significant role, at least in shaping cluster mass and luminosity distributions. Nevertheless it should be noted that, given the relatively large error bars due to small number statistics, weak trends with global star formation properties cannot be ruled out.

We also find no significant trend between the power-law index of the age distributions derived from clusters in the age intervals between 5-500~Myr with \sfrsig. The median fitted slopes of $\approx$-0.8$\pm$0.15 for an ensemble of clusters in our sample of dwarf and irregular galaxies indicate early cluster dissolution at the $\approx$80\% level. While there is evidence for some contamination of cluster samples in the 1-10~Myr age range from clusters older than 100~Myr due to poor fitting resulting from the age-extinction degeneracy, our results show good agreement with those from \cite{whitmore20} whose SED fitting method utilizes \ha~fluxes that effectively break this degeneracy. 

A main focus of this paper is to study the relationship between the fraction of stars in clusters ($\Gamma$) and \sfrsig, where we have examined these properties in different age intervals and have been careful to compare our results matched to the same age intervals. We do not find a significant correlation between $\Gamma$ and \sfrsig\ in any age interval nor when using different SFR tracers.  This could indicate that $\Gamma$ does not significantly change with global star formation properties in dwarf and irregular galaxies or in the \sfrsig\ range covered. However, we cannot rule out a weak trend amongst our sample without better cluster statistics. We also find that the fraction of stars in clusters decreases between 1--10~Myr and 10-100~Myr, with median $\Gamma$ values of $27\pm6$\% and $7\pm2$\%, respectively. This drop in $\Gamma$ from $<10$~Myr to 10-100 Myr is likely due to a combination of various cluster dissolution mechanisms. However, the presence of unbound clusters in the youngest age bin make it difficult to quantify the degree of dissolution. Both of these results at the low \sfrsig\ end are similar to those found by \cite{chandar17} for 8 galaxies covering a larger range of \sfrsig, but which relied on less homogeneous cluster samples. Additional work is needed at the very high \sfrsig\ end to establish if the fraction of stars in bound clusters is higher in those extreme environments in different age intervals.

In future work, it will be interesting to compute $\Gamma$ and \sfrsig~in a series of physical scales spanning a few kpc down to a few 10s of pc to test if $\Gamma$ (and other cluster-host relationships) shows a stronger dependence on local environment. We leave this analysis for a future study.


\section*{Acknowledgements}

Based on observations made with the NASA/ESA Hubble Space Telescope, obtained at the Space Telescope Science Institute, which is operated by the Association of Universities for Research in Astronomy, Inc., under NASA contract NAS 5-26555. These observations are associated with program \# 13364. This research has made use of the NASA/IPAC Extragalactic Database (NED) which is operated by the Jet Propulsion Laboratory, California Institute of Technology, under contract with NASA.  

A.A. acknowledges the support of the Swedish Research Council (Vetenskapsradet) and the Swedish National Space Board (SNSB). M.F. acknowledges support from the European Research Council (ERC) under the European Union's Horizon 2020 research and innovation programme (grant agreement No 757535). 

\section*{Data Availability}
The LEGUS star cluster catalogs are available at https://legus.stsci.edu/ or can be requested by contacting the author. The galaxy properties and $\Gamma$ measurements are available upon request.

\bibliographystyle{texstuff/mn2e}   
\bibliography{texstuff/all}

\appendix

\section{Cluster Distributions for Individual Galaxies} \label{sec:append_lfmfdndt}

\begin{figure*}
  \begin{center}
  \includegraphics[scale=0.27]{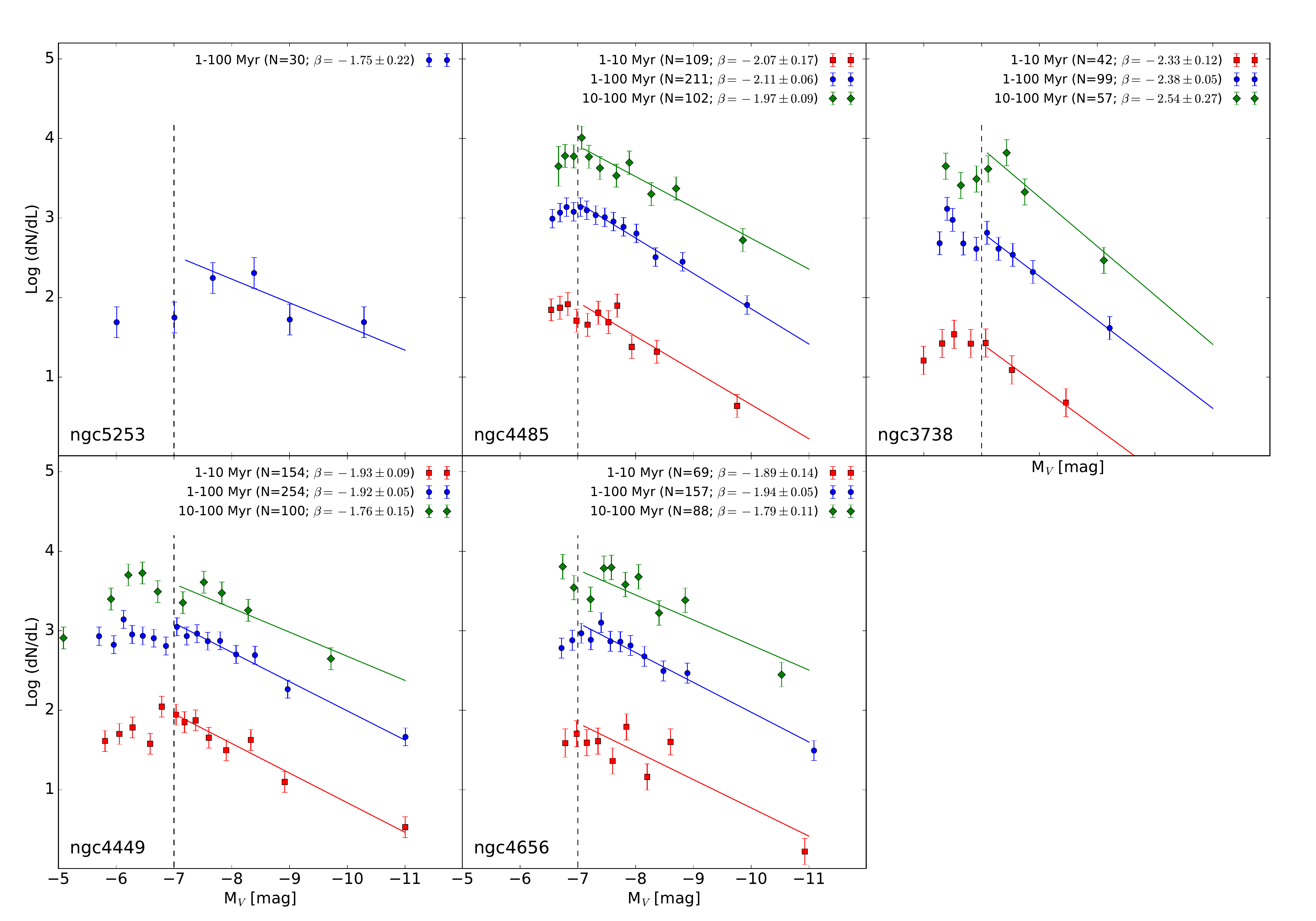}
  \caption{The luminosity functions (LFs) of individual galaxies in the three age ranges studied here (1-10, 1-100, and 10-100~Myr), where there exist a large enough cluster population to provide robust power-law fits (N$\geq$30). The slopes are determined by fitting a power-law to data brighter than the assumed completeness limit of --7~mag. The red square, blue circle, and green diamond symbols represent age ranges of 1-10, 1-100, and 10-100~Myr, respectively. The galaxies are sorted by \sfrsig~using the 10-100~Myr SFHs and the D25$\cap$HST FOV areas (see \S\ref{sec:areatest}).}
   \label{fig:galLF}
   \end{center}
\end{figure*}

\begin{figure*}
  \begin{center}
  \includegraphics[scale=0.27]{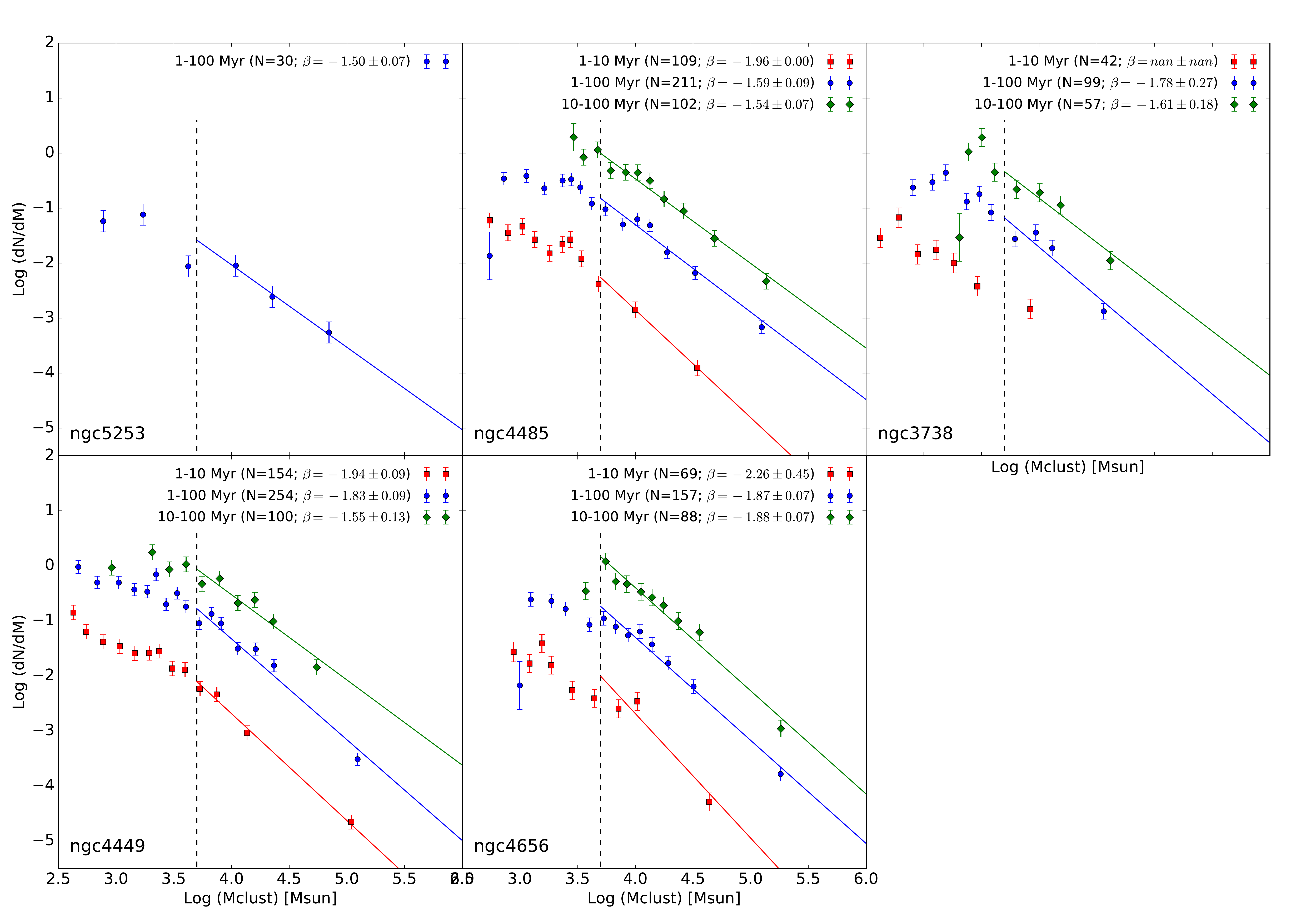}
  \caption{The mass functions (MFs) of individual galaxies where there exist a large enough cluster population to provide robust power-law fits. The slopes are determined by fitting a power-law to data more massive than the assumed completeness limit of Log($M$/M$_{\odot})>$3.7. The layout and symbols are described in Figure~\ref{fig:galLF}}
   \label{fig:galMF}
   \end{center}
\end{figure*}

\begin{figure*}
  \begin{center}
  \includegraphics[scale=0.27]{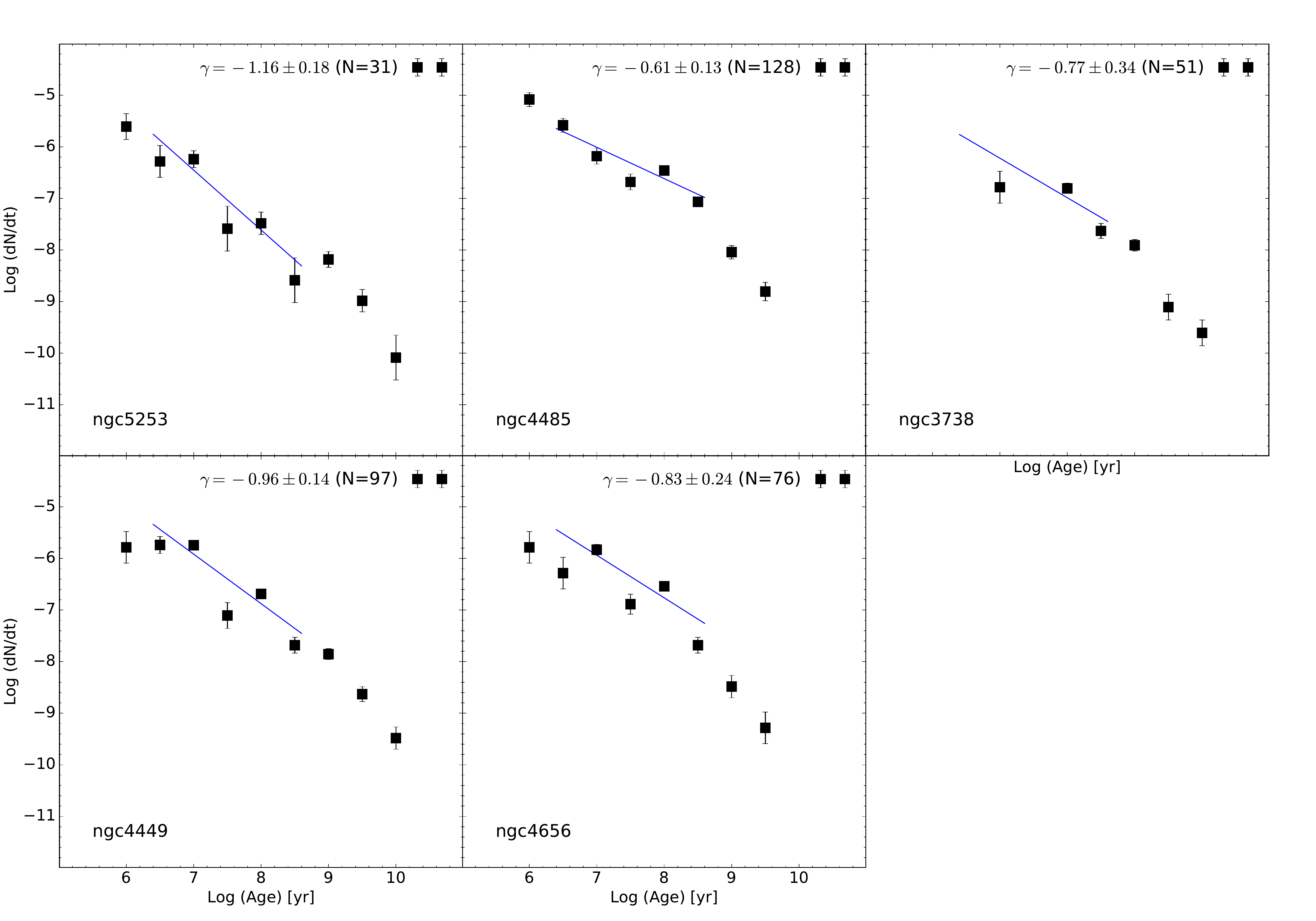}
  \caption{The age distributions of individual galaxies where there exist a large enough cluster population to provide robust power-law fits. The slopes are determined by fitting a power-law to data with a mass cut of Log($M/{\rm M_{\odot}}$)$>$4 and to ages between $6.5 < {\rm Log}(Age) < 8.7$. }
   \label{fig:galdndt}
   \end{center}
\end{figure*}

Here we present the cluster distributions (luminosity, mass, and age) for individual galaxies in the LEGUS dwarf sample. There are only 5 galaxies that show large enough cluster populations for robust power-law fits, and they are presented in Figures~\ref{fig:galLF}--\ref{fig:galdndt}. The power-law fits are performed with the same methods and limits as those described in binned distribution sections of \S\ref{sec:mfresults} and \S\ref{sec:dndtresults}. We note that the galaxies displayed in these figures have been sorted by \sfrsig~using the 10-100~Myr SFHs and the D25$\cap$HST FOV areas.

We find well behaved luminosity and mass distributions, and that the limits to which the fits are applied generally agree with the distribution turnovers. We also find that the slopes are largely consistent with a --2 power-law slope, and there is no evidence of a trend with \sfrsig. However, we note that the range in \sfrsig~is only 0.5~dex (i.e., --1.83 $<$ Log(\sfrsig) $<$ --1.31). Consequently, there may not be enough \sfrsig~range to provide a meaningful trend. 

Figure~\ref{fig:galdndt} shows fairly well behaved age distributions with smooth declines in the clusters numbers over 5--500~Myr and occasionally out to 1~Gyr (e.g., NGC~5253, NGC~3738, and NGC~4449). We find a relatively consistent slope of --0.8$\pm$0.2 and no trend between these slopes and \sfrsig. However, as noted in the previous paragraph, we may not expect to see a trend given the small range of \sfrsig~exhibited by these galaxies. 

\clearpage
\onecolumn 
\section{Literature Data} \label{sec:append_lit}

The data shown in Figures~\ref{fig:MvSFR} and Figure~\ref{fig:gammasig} came from a variety of studies going back twenty years and are tabulated in Table~\ref{tab:lit}. The age of the clusters and the galaxy's SFRs rates are provided when the study gave explicit values. However, many of the older studies only reported either that the clusters were young or that the SFRs used were recent (i.e., less than a few hundred Myr); the age ranges of these studies are denoted with ellipses in the table. The compilation of data in this table is unique as it combines: the brightest cluster, galaxy star formation rates, the fraction of stars in clusters, and the age ranges of these data. 


\begin{longtable}{llrrrcc}
\caption{The literature data used in Figures~\ref{fig:MvSFR} and Figure~\ref{fig:gammasig} are tabulated here, and include data from: \citet{adamo20},
\citet{Randriamanakoto19},
\citet{messa18a},
\citet{chandar17},
\citet{johnson16},
\citet{Hollyhead16},
\citet{lim15},
\citet{adamo15},
\citet{whitmore14a},
\citet{Ryon14},
\citet{baumgardt03},
\citet{cook12},
\citet{silvavilla11},
\citet{Pasquali11},
\citet{annibali11},
\citet{adamo11c},
\citet{goddard10},
\citet{Annibali09},
\citet{bastian08},
\citet{larsen02},
\citet{billett02},
\citet{larsenrichtler00},
\citet{Johnson00}.
 } \\\label{tab:lit}
\\ 
\hline
\hline
Reference & Galaxy & M$^{brightest}_{V}$ & Log(SFR)              & Log(\sfrsig)                   & $\Gamma$  & Age   \\ 
          &        & (mag)               & ($M_{\star} yr^{-1}$) & ($M_{\star} yr^{-1} kpc^{-2}$) & (\%)      & (Myr) \\ 
\hline
\endfirsthead

\hline
Reference & Galaxy & M$^{brightest}_{V}$ & Log(SFR)              & Log(\sfrsig)                   & $\Gamma$  & Age   \\ 
          &        & (mag)               & ($M_{\star} yr^{-1}$) & ($M_{\star} yr^{-1} kpc^{-2}$) & (\%)      & (Myr) \\ 
\hline
\endhead

\hline
\endfoot

\hline
\endlastfoot
Adamo+20 &&&&&&\\ 
         & NGC6052 & \ldots & 1.18 & -0.89 & 40.80 $\pm$ 2.0 & 1-10 \\ 
         & NGC4194 & \ldots & 1.13 & -0.34 & 69.80 $\pm$ 4.9 & 1-10 \\ 
         & NGC3690B & \ldots & 1.27 & -0.40 & 59.20 $\pm$ 6.6 & 1-10 \\ 
         & NGC3690A & \ldots & 1.46 & -0.55 & 32.00 $\pm$ 4.7 & 1-10 \\ 
         & NGC3690 & \ldots & 1.68 & -0.51 & 41.40 $\pm$ 3.8 & 1-10 \\ 
         & NGC34 & \ldots & 0.76 & -1.15 & 38.90 $\pm$ 1.7 & 1-10 \\ 
         & NGC3256 & \ldots & 1.65 & -0.24 & 54.10 $\pm$ 3.2 & 1-10 \\ 
         & NGC1614 & \ldots & 1.44 & -0.47 & 83.10 $\pm$ 15.2 & 1-10 \\ 
Randriamanakoto+19 &&&&&&\\ 
         & ARP299 & -15.20 & 1.94 & -0.96 & 16.00 $\pm$ 4.4 & 10-100 \\ 
Messa+18a &&&&&&\\ 
         & M51 & \ldots & 0.21 & -1.86 & 18.60 $\pm$ 2.4 & 10-100 \\ 
         & M51 & \ldots & 0.21 & -1.86 & 19.60 $\pm$ 2.5 & 1-100 \\ 
         & M51 & \ldots & 0.16 & -1.88 & 32.40 $\pm$ 12.1 & 1-10 \\ 
Chandar+17 &&&&&&\\ 
         & SMC & \ldots & -1.22 & -3.04 & 3.00 $\pm$ 2.0 & 10-100 \\ 
         & SMC & \ldots & -1.22 & -3.04 & 36.00 $\pm$ 23.0 & 1-10 \\ 
         & NGC4449 & \ldots & -0.46 & -1.89 & 3.00 $\pm$ 2.0 & 10-100 \\ 
         & NGC4449 & \ldots & -0.46 & -1.89 & 28.00 $\pm$ 19.0 & 1-10 \\ 
         & NGC4214 & \ldots & -0.96 & -1.66 & 2.00 $\pm$ 1.0 & 10-100 \\ 
         & NGC4214 & \ldots & -0.96 & -1.66 & 8.00 $\pm$ 5.0 & 1-10 \\ 
         & NGC3256 & \ldots & 1.70 & 0.00 & 3.00 $\pm$ 2.0 & 10-100 \\ 
         & NGC3256 & \ldots & 1.70 & 0.00 & 26.00 $\pm$ 17.0 & 1-10 \\ 
         & M83 & \ldots & 0.42 & -1.89 & 10.00 $\pm$ 5.0 & 10-100 \\ 
         & M83 & \ldots & 0.42 & -1.89 & 12.00 $\pm$ 8.0 & 1-10 \\ 
         & M51 & \ldots & 0.51 & -1.80 & 5.00 $\pm$ 3.0 & 10-100 \\ 
         & M51 & \ldots & 0.51 & -1.80 & 30.00 $\pm$ 18.0 & 1-10 \\ 
         & LMC & \ldots & -0.60 & -2.66 & 5.00 $\pm$ 3.0 & 10-100 \\ 
         & LMC & \ldots & -0.60 & -2.66 & 27.00 $\pm$ 18.0 & 1-10 \\ 
         & Antennae & \ldots & 1.30 & -0.76 & 6.00 $\pm$ 4.0 & 10-100 \\ 
         & Antennae & \ldots & 1.30 & -0.76 & 22.00 $\pm$ 14.0 & 1-10 \\ 
Johnson+16 &&&&&&\\ 
         & M31 & \ldots & \ldots & -2.63 & 5.90 $\pm$ 0.3 & 10-100 \\ 
Hollyhead+16 &&&&&&\\ 
         & NGC1566 & -13.40 & 0.63 & -1.48 & 8.80 $\pm$ 1.1 & 1-10 \\ 
Lim\&Lee+15 &&&&&&\\ 
         & IC10 & -10.40 & -1.15 & -1.52 & 4.20 & 1-10 \\ 
Adamo+15 &&&&&&\\ 
         & M83 & -11.60 & -0.09 & -1.89 & 18.20 $\pm$ 3.0 & 1-10 \\ 
Whitmore+14 &&&&&&\\ 
         & NGC7793 & -9.65 & -1.18 & \ldots & \ldots & 1-100 \\ 
         & NGC6503 & -10.51 & -1.16 & \ldots & \ldots & 1-100 \\ 
         & NGC628/M74 & -11.84 & -0.64 & \ldots & \ldots & 1-100 \\ 
         & NGC6217 & -13.28 & \ldots & \ldots & \ldots & 1-100 \\ 
         & NGC5457(field2)/M101 & -11.57 & -0.69 & \ldots & \ldots & 1-100 \\ 
         & NGC5457(field1)/M101 & -11.38 & -0.59 & \ldots & \ldots & 1-100 \\ 
         & NGC5236(field2)/M83 & -10.06 & -1.01 & \ldots & \ldots & 1-100 \\ 
         & NGC5236(field1)/M83 & -11.74 & -0.54 & \ldots & \ldots & 1-100 \\ 
         & NGC5055/M63 & -9.61 & -1.63 & \ldots & \ldots & 1-100 \\ 
         & NGC4736 & -10.44 & -1.41 & \ldots & \ldots & 1-100 \\ 
         & NGC45 & -10.83 & -0.93 & \ldots & \ldots & 1-100 \\ 
         & NGC4395 & -9.79 & -1.14 & \ldots & \ldots & 1-100 \\ 
         & NGC4394 & -10.25 & -0.85 & \ldots & \ldots & 1-100 \\ 
         & NGC4258 & -10.64 & -1.01 & \ldots & \ldots & 1-100 \\ 
         & NGC406 & -11.75 & -0.53 & \ldots & \ldots & 1-100 \\ 
         & NGC4038 & -15.25 & 0.39 & \ldots & \ldots & 1-100 \\ 
         & NGC3627/M66 & -11.97 & -0.43 & \ldots & \ldots & 1-100 \\ 
         & NGC2397 & -13.47 & \ldots & \ldots & \ldots & 1-100 \\ 
         & NGC1483 & -10.01 & -0.96 & \ldots & \ldots & 1-100 \\ 
         & NGC1313 & -10.98 & -0.66 & \ldots & \ldots & 1-100 \\ 
         & NGC1309 & -13.80 & 0.23 & \ldots & \ldots & 1-100 \\ 
         & NGC1300(field1) & -11.52 & -0.49 & \ldots & \ldots & 1-100 \\ 
         & NGC1300(field2) & -11.00 & -0.52 & \ldots & \ldots & 1-100 \\ 
Ryon+14 &&&&&&\\ 
         & NGC2997 & \ldots & -0.66 & -2.02 & 10.00 $\pm$ 3.0 & 1-100 \\ 
Baumgardt+13 &&&&&&\\ 
         & LMC & -10.95 & -0.54 & -2.43 & 15.00 & 10-100 \\ 
Cook+12 &&&&&&\\ 
         & UGCA292 & -7.46 & -2.47 & -2.80 & 10.63 $\pm$ 2.8 & 1-100 \\ 
         & UGCA292 & \ldots & -2.29 & -2.62 & \ldots & 1-10 \\ 
         & UGCA276 & \ldots & -4.70 & -5.27 & \ldots & 1-100 \\ 
         & UGCA276 & \ldots & \ldots & \ldots & \ldots & 1-10 \\ 
         & UGCA133 & \ldots & -4.73 & -5.49 & \ldots & 1-100 \\ 
         & UGCA133 & \ldots & \ldots & \ldots & \ldots & 1-10 \\ 
         & UGC9240 & -5.90 & -2.05 & -2.82 & 3.34 $\pm$ 0.6 & 1-100 \\ 
         & UGC9240 & \ldots & -2.28 & -3.04 & \ldots & 1-10 \\ 
         & UGC9128 & -5.45 & -2.89 & -2.90 & 5.12 $\pm$ 1.1 & 1-100 \\ 
         & UGC9128 & \ldots & -3.55 & -3.56 & \ldots & 1-10 \\ 
         & UGC8833 & \ldots & -2.81 & -3.19 & \ldots & 1-100 \\ 
         & UGC8833 & \ldots & -2.60 & -2.99 & \ldots & 1-10 \\ 
         & UGC8760 & -7.37 & -2.37 & -3.03 & 1.56 $\pm$ 0.3 & 1-100 \\ 
         & UGC8760 & -7.37 & -2.52 & -3.18 & 35.33 $\pm$ 8.3 & 1-10 \\ 
         & UGC8651 & \ldots & -2.47 & -3.16 & \ldots & 1-100 \\ 
         & UGC8651 & \ldots & -3.08 & -3.77 & \ldots & 1-10 \\ 
         & UGC8508 & \ldots & -2.56 & -2.93 & \ldots & 1-100 \\ 
         & UGC8508 & \ldots & -2.18 & -2.56 & \ldots & 1-10 \\ 
         & UGC8201 & -9.16 & -1.24 & -2.43 & 2.97 $\pm$ 0.4 & 1-100 \\ 
         & UGC8201 & \ldots & -1.50 & -2.69 & \ldots & 1-10 \\ 
         & UGC5692 & -9.44 & -2.89 & -4.13 & 5.23 $\pm$ 1.2 & 1-100 \\ 
         & UGC5692 & -9.44 & -2.13 & -3.37 & 14.63 $\pm$ 3.1 & 1-10 \\ 
         & UGC5442 & \ldots & \ldots & \ldots & \ldots & 1-100 \\ 
         & UGC5442 & \ldots & \ldots & \ldots & \ldots & 1-10 \\ 
         & UGC5428 & \ldots & \ldots & \ldots & \ldots & 1-100 \\ 
         & UGC5428 & \ldots & \ldots & \ldots & \ldots & 1-10 \\ 
         & UGC5336 & -8.59 & -1.57 & -2.60 & 1.90 $\pm$ 0.2 & 1-100 \\ 
         & UGC5336 & -8.59 & -2.43 & -3.46 & 73.28 $\pm$ 37.6 & 1-10 \\ 
         & UGC5139 & \ldots & -1.84 & -3.06 & \ldots & 1-100 \\ 
         & UGC5139 & \ldots & -1.76 & -2.97 & \ldots & 1-10 \\ 
         & UGC4459 & -7.90 & -2.19 & -2.74 & 4.26 $\pm$ 0.6 & 1-100 \\ 
         & UGC4459 & -7.90 & -2.49 & -3.05 & 136.76 $\pm$ 19.6 & 1-10 \\ 
         & UGC4305 & -8.88 & -1.20 & -2.93 & 2.66 $\pm$ 0.3 & 1-100 \\ 
         & UGC4305 & -8.88 & -0.94 & -2.68 & 13.29 $\pm$ 0.8 & 1-10 \\ 
         & NGC4163 & \ldots & -2.59 & -3.31 & \ldots & 1-100 \\ 
         & NGC4163 & \ldots & -2.00 & -2.73 & \ldots & 1-10 \\ 
         & NGC404 & \ldots & \ldots & \ldots & \ldots & 1-100 \\ 
         & NGC404 & \ldots & \ldots & \ldots & \ldots & 1-10 \\ 
         & NGC3741 & \ldots & -2.37 & -3.00 & \ldots & 1-100 \\ 
         & NGC3741 & \ldots & -1.84 & -2.46 & \ldots & 1-10 \\ 
         & NGC3077 & \ldots & \ldots & \ldots & \ldots & 1-100 \\ 
         & NGC3077 & \ldots & \ldots & \ldots & \ldots & 1-10 \\ 
         & NGC2366 & -8.52 & -1.20 & -2.77 & 0.51 $\pm$ 0.1 & 1-100 \\ 
         & NGC2366 & -8.52 & -1.03 & -2.60 & 1.44 $\pm$ 0.4 & 1-10 \\ 
         & M81DA & \ldots & -2.78 & -2.90 & \ldots & 1-100 \\ 
         & M81DA & \ldots & -2.68 & -2.81 & \ldots & 1-10 \\ 
         & KKR3 & \ldots & -3.62 & -2.99 & \ldots & 1-100 \\ 
         & KKR3 & \ldots & \ldots & \ldots & \ldots & 1-10 \\ 
         & KKH98 & \ldots & -3.30 & -3.38 & \ldots & 1-100 \\ 
         & KKH98 & \ldots & \ldots & \ldots & \ldots & 1-10 \\ 
         & KKH37 & \ldots & -3.33 & -3.54 & \ldots & 1-100 \\ 
         & KKH37 & \ldots & \ldots & \ldots & \ldots & 1-10 \\ 
         & KK77 & \ldots & -4.87 & -5.13 & \ldots & 1-100 \\ 
         & KK77 & \ldots & \ldots & \ldots & \ldots & 1-10 \\ 
         & KDG73 & \ldots & -3.09 & -3.67 & \ldots & 1-100 \\ 
         & KDG73 & \ldots & \ldots & \ldots & \ldots & 1-10 \\ 
         & KDG61 & \ldots & -3.99 & -4.75 & \ldots & 1-100 \\ 
         & KDG61 & \ldots & -3.33 & -4.09 & \ldots & 1-10 \\ 
         & KDG2 & \ldots & -4.64 & -5.35 & \ldots & 1-100 \\ 
         & KDG2 & \ldots & \ldots & \ldots & \ldots & 1-10 \\ 
         & IKN & \ldots & -3.79 & -4.61 & \ldots & 1-100 \\ 
         & IKN & \ldots & \ldots & \ldots & \ldots & 1-10 \\ 
         & IC2574 & -9.12 & -1.09 & -3.14 & 2.44 $\pm$ 0.3 & 1-100 \\ 
         & IC2574 & -9.12 & -0.97 & -3.03 & 8.13 $\pm$ 1.4 & 1-10 \\ 
         & HS117 & \ldots & -5.05 & -5.91 & \ldots & 1-100 \\ 
         & HS117 & \ldots & -3.85 & -4.71 & \ldots & 1-10 \\ 
         & GR8 & \ldots & \ldots & \ldots & \ldots & 1-100 \\ 
         & GR8 & \ldots & \ldots & \ldots & \ldots & 1-10 \\ 
         & GARLAND & \ldots & \ldots & \ldots & \ldots & 1-100 \\ 
         & GARLAND & \ldots & \ldots & \ldots & \ldots & 1-10 \\ 
         & FM1 & \ldots & -5.22 & -5.47 & \ldots & 1-100 \\ 
         & FM1 & \ldots & -5.44 & -5.70 & \ldots & 1-10 \\ 
         & DDO78 & \ldots & -4.66 & -5.35 & \ldots & 1-100 \\ 
         & DDO78 & \ldots & -5.28 & -5.97 & \ldots & 1-10 \\ 
         & BK3N & \ldots & -2.80 & -2.45 & \ldots & 1-100 \\ 
         & BK3N & \ldots & -4.87 & -4.51 & \ldots & 1-10 \\ 
         & ARPSLOOP & \ldots & -2.60 & -3.28 & \ldots & 1-100 \\ 
         & ARPSLOOP & \ldots & \ldots & \ldots & \ldots & 1-10 \\ 
SVL+11 &&&&&&\\ 
         & NGC7793 & \ldots & -0.82 & -2.19 & 9.80 & 10-100 \\ 
         & NGC5236 & \ldots & -0.41 & -1.87 & 9.80 & 10-100 \\ 
         & NGC45 & \ldots & -1.30 & -2.99 & 17.30 & 10-100 \\ 
         & NGC4395 & \ldots & -0.77 & -2.33 & 2.60 & 10-100 \\ 
         & NGC1313 & \ldots & -0.17 & -1.95 & 9.00 & 10-100 \\ 
Pasquali+11 &&&&&&\\ 
         & NGC1569 & -13.90 & -0.44 & -1.52 & 13.90 $\pm$ 0.8 & 1-10 \\ 
Annibali+11 &&&&&&\\ 
         & NGC4449 & \ldots & 0.00 & -1.40 & 10.00 & 1-10 \\ 
Adamo+11 &&&&&&\\ 
         & SBS0335 & -14.28 & 0.11 & -0.02 & 49.00 $\pm$ 15.0 & 1-10 \\ 
         & MRK930 & -15.17 & 0.73 & -0.23 & 25.00 $\pm$ 10.0 & 1-10 \\ 
         & HARO11 & -16.16 & 1.34 & 0.33 & 50.00 $\pm$ 14.0 & 1-10 \\ 
         & ESO338 & -15.50 & 0.51 & 0.19 & 50.00 $\pm$ 10.0 & 1-10 \\ 
         & ESO185 & -14.55 & 0.81 & -0.28 & 26.00 $\pm$ 5.0 & 1-10 \\ 
Goddard+10 &&&&&&\\ 
         & SMC & \ldots & -1.37 & -3.14 & 4.20 $\pm$ 0.2 & 10-100 \\ 
         & NGC6946 & \ldots & -0.76 & -2.34 & 12.50 $\pm$ 2.0 & 1-10 \\ 
         & NGC5236 & \ldots & -0.41 & -0.26 & 26.70 $\pm$ 5.0 & 1-10 \\ 
         & NGC3256 & \ldots & 1.66 & -0.21 & 22.90 $\pm$ 8.0 & 1-10 \\ 
         & NGC1569 & \ldots & -0.44 & -1.55 & 13.90 $\pm$ 0.8 & 1-10 \\ 
         & MW & \ldots & -0.82 & -1.92 & 7.00 $\pm$ 5.0 & 1-10 \\ 
         & LMC & \ldots & -0.92 & -2.82 & 5.80 $\pm$ 0.5 & 10-100 \\ 
Annibali+09 &&&&&&\\ 
         & NGC1705 & -13.80 & -0.51 & -1.34 & \ldots & 1-10 \\ 
Bastian+08 &&&&&&\\ 
         & NGC7673c & -14.70 & 0.69 & \ldots & \ldots & \ldots \\ 
         & NGC7252a & -13.40 & 0.73 & \ldots & \ldots & \ldots \\ 
         & NGC7252 & -17.30 & 2.76 & \ldots & \ldots & \ldots \\ 
         & NGC7252 & -18.90 & 3.62 & \ldots & \ldots & \ldots \\ 
         & NGC6745 & -15.00 & 1.09 & \ldots & \ldots & \ldots \\ 
         & NGC6240b & -16.40 & 2.15 & \ldots & \ldots & \ldots \\ 
         & NGC3921 & -15.30 & 1.70 & \ldots & \ldots & \ldots \\ 
         & NGC3610 & -16.50 & 2.34 & \ldots & \ldots & \ldots \\ 
         & NGC3597 & -13.30 & 1.03 & \ldots & \ldots & \ldots \\ 
         & NGC3597 & -16.40 & 1.76 & \ldots & \ldots & \ldots \\ 
         & NGC34 & -17.30 & 2.78 & \ldots & \ldots & \ldots \\ 
         & NGC3256 & -15.70 & 1.66 & \ldots & \ldots & \ldots \\ 
         & NGC2623 & -14.50 & 1.71 & \ldots & \ldots & \ldots \\ 
         & NGC2207 & -13.60 & 0.34 & \ldots & \ldots & \ldots \\ 
         & NGC1700 & -15.80 & 1.96 & \ldots & \ldots & \ldots \\ 
         & NGC1533Assn1 & -7.17 & -3.43 & \ldots & \ldots & \ldots \\ 
         & NGC1533Assn2 & -5.71 & -3.60 & \ldots & \ldots & \ldots \\ 
         & NGC1533Assn5 & -6.16 & -3.74 & \ldots & \ldots & \ldots \\ 
         & NGC1316 & -17.60 & 2.93 & \ldots & \ldots & \ldots \\ 
         & NGC1275 & -15.30 & 1.09 & \ldots & \ldots & \ldots \\ 
         & NGC1140 & -14.80 & -0.10 & \ldots & \ldots & \ldots \\ 
         & M82(A1) & -14.80 & 0.85 & \ldots & \ldots & \ldots \\ 
         & IRAS19115-2124d & -16.80 & 2.28 & \ldots & \ldots & \ldots \\ 
         & ESO0338-IG04 & -15.50 & 0.51 & \ldots & \ldots & \ldots \\ 
Larsen+02 &&&&&&\\ 
         & NGC7793 & -10.40 & -0.85 & -2.67 & \ldots & \ldots \\ 
         & NGC7424 & -11.40 & -0.76 & -3.74 & \ldots & \ldots \\ 
         & NGC6946a & -13.00 & 0.40 & -2.34 & \ldots & \ldots \\ 
         & NGC6744a & -11.00 & -0.37 & -3.21 & \ldots & \ldots \\ 
         & NGC628a & -11.30 & -0.00 & -2.73 & \ldots & \ldots \\ 
         & NGC5585 & -10.80 & -1.47 & -3.49 & \ldots & \ldots \\ 
         & NGC5236a & -11.70 & 0.36 & -1.86 & \ldots & \ldots \\ 
         & NGC5204 & -9.60 & -1.50 & -3.08 & \ldots & \ldots \\ 
         & NGC5194 & -12.80 & 0.68 & -2.09 & \ldots & \ldots \\ 
         & NGC5055 & -11.40 & 0.18 & -2.53 & \ldots & \ldots \\ 
         & NGC45 & -8.80 & -1.60 & -3.64 & \ldots & \ldots \\ 
         & NGC4395 & -9.10 & -1.30 & -3.60 & \ldots & \ldots \\ 
         & NGC4258 & -12.60 & -0.00 & -3.15 & \ldots & \ldots \\ 
         & NGC3621 & -11.90 & -0.06 & -2.78 & \ldots & \ldots \\ 
         & NGC3521 & -11.50 & 0.14 & -2.45 & \ldots & \ldots \\ 
         & NGC3184a & -10.60 & -0.40 & -2.76 & \ldots & \ldots \\ 
         & NGC300 & -9.90 & -1.11 & -3.31 & \ldots & \ldots \\ 
         & NGC2997 & -12.90 & 0.27 & -2.51 & \ldots & \ldots \\ 
         & NGC2835 & -10.90 & -1.04 & -3.14 & \ldots & \ldots \\ 
         & NGC247 & -10.20 & -1.44 & -3.74 & \ldots & \ldots \\ 
         & NGC2403 & -9.90 & -0.47 & -3.01 & \ldots & \ldots \\ 
         & NGC1313a & -12.10 & -0.37 & -2.39 & \ldots & \ldots \\ 
         & NGC1156 & -11.10 & -0.73 & -2.51 & \ldots & \ldots \\ 
         & IC2574 & -10.50 & -1.69 & -3.77 & \ldots & \ldots \\ 
BHE+02 &&&&&&\\ 
         & SextansA & -7.12 & -2.44 & -2.64 & \ldots & \ldots \\ 
         & NGC4214 & -12.04 & -1.10 & -2.42 & \ldots & \ldots \\ 
         & NGC2366 & -9.51 & -1.43 & -2.71 & \ldots & \ldots \\ 
         & DDO50 & -7.91 & -1.97 & -2.90 & \ldots & \ldots \\ 
         & DDO168 & -7.58 & -2.36 & -3.07 & \ldots & \ldots \\ 
         & DDO165 & -8.34 & -3.27 & -3.74 & \ldots & \ldots \\ 
LR+00 &&&&&&\\ 
         & NGC5253 & -11.10 & -0.67 & -2.14 & \ldots & \ldots \\ 
         & NGC1741 & -15.00 & 0.69 & -1.89 & \ldots & \ldots \\ 
         & NGC1705 & -13.70 & -1.89 & -2.67 & \ldots & \ldots \\ 
         & NGC1569 & -13.90 & -0.91 & -2.03 & \ldots & \ldots \\ 
         & LMC & -10.00 & -0.92 & -2.82 & \ldots & \ldots \\ 
         & IC1613 & -5.80 & -3.35 & -4.30 & \ldots & \ldots \\ 
Johnson+00 &&&&&&\\ 
         & He2-10 & -12.50 & -0.70 & \ldots & \ldots & 1-10 \\ 

\hline

\end{longtable}

\twocolumn 

\end{document}